\documentclass[twocolumn,twocolappendix,longauth]{aa}
\usepackage{graphicx}
\usepackage{natbib}
\usepackage[switch]{lineno}
\usepackage[utf8]{inputenc}
\usepackage{xcolor}
\usepackage{booktabs}
\usepackage{multirow}

\usepackage{float}
\usepackage{txfonts}
\usepackage{hyperref}
\hypersetup{hidelinks}

\begin{document}

   \title{Complex Organic Molecules in Protostars with ALMA Spectral Surveys (COMPASS)}
   \subtitle{VI. Discovery of a class I methanol maser transition and its association with acetaldehyde}

   \author{Jae-Hong Jeong\inst{\ref{inst1}} \and Jeong-Eun Lee\inst{\ref{inst1}} \and Jes K. J{\o}rgensen\inst{\ref{inst2}} \and Audrey Coutens\inst{\ref{inst3}} \and Maria N. Drozdovskaya\inst{\ref{inst4},\ref{inst18}} \and Adele L. Plunkett\inst{\ref{inst5}} \and Arnaud Belloche\inst{\ref{inst6}} \and S{\'e}bastien Maret\inst{\ref{inst7}} \and Hyeong-Sik Yun\inst{\ref{inst1},\ref{inst17}} \and Fernando Cruz-S\'aenz de Miera\inst{\ref{inst3}} \and Daniel Harsono\inst{\ref{inst16}} \and Merel L. R. van 't Hoff\inst{\ref{inst8}} \and Chul-Hwan Kim\inst{\ref{inst1}} \and \'Agnes K\'osp\'al\inst{\ref{inst9},\ref{inst10},\ref{inst11}} \and  Brett A. McGuire\inst{\ref{inst12}} \and  Pooneh Nazari\inst{\ref{inst13}} \and Silvia Spezzano\inst{\ref{inst14}} \and Yao-Lun Yang\inst{\ref{inst15}} \and Shaoshan Zeng\inst{\ref{inst15}}
          }

   \institute{
        Department of Physics and Astronomy, Seoul National University, 1 Gwanak-ro, Gwanak-gu, Seoul 08826, Republic of Korea\\\email{lee.jeongeun@snu.ac.kr}\label{inst1}
         \and
        Niels Bohr Institute, University of Copenhagen, Jagtvej 155A, DK-2200 Copenhagen N, Denmark\label{inst2}
         \and
        Institut de Recherche en Astrophysique et Plan\'etologie, Universit\'e de Toulouse, CNRS, CNES, 9 av. du Colonel Roche, 31028 Toulouse Cedex 4, France\label{inst3}
        \and
        Physikalisch-Meteorologisches Observatorium Davos und Weltstrahlungszentrum (PMOD/WRC), Dorfstrasse 33, 7260 Davos Dorf, Switzerland\label{inst4}
        \and
        Department of Chemistry, Biochemistry and Pharmaceutical Sciences (DCBP), Universit{\"a}t Bern, Freiestrasse 3, 3012 Bern, Switzerland\label{inst18}
        \and
        National Radio Astronomy Observatory, 520 Edgemont Road, Charlottesville, VA 22903, USA\label{inst5}
        \and
        Max-Planck-Institut f{\"u}r Radioastronomie, Auf dem H{\"u}gel 69, 53121 Bonn, Germany\label{inst6}
        \and
        Institut de Plan\'etologie et d'Astrophysique de Grenoble, Univ. Grenoble Alpes, CNRS, IPAG, 38000 Grenoble, France\label{inst7}
        \and
        Korea Astronomy and Space Science Institute, 776 Daedeok-daero, Yuseong-gu, Daejeon 34055, Republic of Korea\label{inst17}
        \and
        Institute of Astronomy, Department of Physics, National Tsinghua University, No. 101, Section 2, Kuang-Fu Road, Hsinchu 30013, Taiwan, R.O.C.\label{inst16}
        \and
        Department of Physics and Astronomy, Purdue University, 525 Northwestern Ave, West Lafayette, IN 47907, USA\label{inst8}
        \and
        Konkoly Observatory, HUN-REN Research Centre for Astronomy and Earth Sciences, MTA Centre of Excellence, Konkoly-Thege Mikl\'os \'ut 15-17, 1121 Budapest, Hungary\label{inst9}
        \and
        Institute of Physics and Astronomy, ELTE E\"otv\"os Lor\'and University, P\'azm\'any P\'eter s\'et\'any 1/A, 1117 Budapest, Hungary\label{inst10}
        \and
        Max-Planck-Institut f\"ur Astronomie, K\"onigstuhl 17, 69117 Heidelberg, Germany\label{inst11}
        \and
        Department of Chemistry, Massachusetts Institute of Technology, 77 Massachusetts Ave, Cambridge, MA 02139, USA\label{inst12}
        \and
        European Southern Observatory, Karl-Schwarzschild-Strasse 2, 85748 Garching, Germany\label{inst13}
        \and
        Max-Planck-Institut f{\"u}r extraterrestrische Physik, Giessenbachstrasse 1, 85748 Garching bei M{\"u}nchen, Germany\label{inst14}
        \and
        Star and Planet Formation Laboratory, RIKEN Pioneering Research Institute, Wako-shi, Saitama, 351-0198, Japan\label{inst15}
        }

   \date{Received December 2025; accepted July 2026}

  \abstract
{Methanol (CH$_3$OH) masers that are collisionally pumped, referred to as `class I', commonly trace shock interfaces in star-forming regions, but their incidence in embedded low-mass protostars remains poorly quantified.}
{We aim to (i) establish the occurrence, spatial origin, and excitation of a newly detected class~I CH$_3$OH maser at 302.913~GHz ($12_{0,12}$–$11_{1,11}$A, $E_{\rm u}=180.9$~K) towards low-mass protostars and (ii) test whether the environments that pump this maser exhibit signatures of shock-driven complex chemistry.}
{We utilised data from the COMPASS ALMA Large Program to search for maser activity in the 302.9~GHz line in a sample of 11 low-mass protostellar sources. The COMPASS programme provides unbiased surveys of these sources between 279.0 and 311.7~GHz covering several transitions of CH$_3$OH as well as other complex organic molecules (COMs) at typical angular resolutions of $0\farcs3$-$0\farcs5$. Measuring departures from local thermodynamic equilibrium (LTE) in the 302.9~GHz CH$_3$OH excitation via rotation diagram analysis robustly revealed masing.}
{The 302.9~GHz CH$_3$OH line is detected as a non-thermal excess (maser) in seven out of 11 COMPASS sources, with peak brightness temperatures ranging from a few to a thousand kelvin. One of these detections is towards HOPS 108, which hosts the brightest maser, and it is likely associated with the powerful outflow driven by the nearby intermediate-mass protostar HOPS 370. Excluding this case, the maser is detected in six of the ten low-mass protostars (60$\%$). All six low-mass protostars in which maser emission is detected are Class~0 sources interacting with dense ambient material, whereas non-detections correspond to three Class~I objects and one isolated Class~0 core. This indicates a preference for early evolutionary stages and environments rich in gas for outflow–ambient medium interactions. The maser emission is spatially coincident with thermal CH$_3$OH emission along the outflow cavity walls and bow shocks, but it has more compact knot-like morphologies. Lastly, among the surveyed COMs, only acetaldehyde (CH$_3$CHO) closely traces the shocked morphology of the 302.9~GHz CH$_3$OH maser. However, this CH$_3$CHO feature is observed only towards the brightest maser in the region around HOPS 108, which is influenced by the intermediate-mass protostellar outflow.}
{Our study reveals that class I CH$_3$OH masers in low-mass protostellar systems are more prevalent than previously recognised and preferentially occur in younger, more embedded systems. Although this has only been observed in an intermediate-mass protostellar outflow, CH$_3$CHO is enhanced in a manner consistent with shock-driven chemistry along the outflow cavities. Testing whether a similar relationship applies in the low-mass regime will require more sensitive CH$_3$CHO observations towards regions where the class I CH$_3$OH masers are detected.}

    \keywords{Masers -- Astrochemistry -- Stars: protostars -- ISM: molecules -- ISM: jets and outflows -- Shock waves}
    \titlerunning{COMPASS CH$_3$OH maser}
    \authorrunning{Jeong, J.-H. et al.}\maketitle

\section{Introduction}
Methanol (CH$_{3}$OH) is the simplest complex organic molecule (COM; i.e. a carbon-bearing molecule that consists of six or more atoms; \citealt{Herbst2009}), and it is formed mostly by the hydrogenation of a CO molecule in the icy mantles of cold interstellar dust grains \citep{Tielens1982, Watanabe2002, Fuchs2009}. It is abundant in star-forming regions and is a key molecule during protostellar evolution, from astrochemical and physical perspectives. Importantly, methanol acts as a parent molecule to make other COMs \citep{Garrod2008, oberg2009}.\par

In general, saturated COMs, including CH$_3$OH, are thermally desorbed from icy dust grains and detected in the gas confined to the warm inner envelopes of protostars. In the case of low-mass protostars, such warm inner envelope regions are known as hot corinos \citep{Ceccarelli2004}. However, non-thermal desorption can give rise to more spatially extended gaseous COMs in star-forming regions (e.g. \citealt{Chengalur2003, Arce2007, oberg2011, Codella2015, Palau2017, Simone2020, Busch2024, Garcia2024, Hsu2024, SHLee2024}). \par

As one of the most abundant COMs, gas-phase CH$_3$OH exhibits numerous rotational transitions observable in various environments. This makes it a particularly powerful tracer of extended structures of gaseous COMs under diverse physical conditions. In addition, some of the CH$_3$OH transitions show strong maser characteristics. Because masers are excited under specific conditions, they serve as sensitive probes of physical and dynamical conditions, enabling detailed investigations of the processes associated with star formation \citep{Pestalozzi2005, Ellingsen2006, Breen2010, Baek2023, Hunter2026}.\par

The condition for masing is a population inversion in the energy levels, which requires a non-thermal source of excitation (pumping). There are two types of CH$_{3}$OH masers \citep{Batrla1987, Menten1991}, and they are classified according to their pumping mechanisms, with class I being collisionally pumped \citep{Plambeck1990, Kurtz2004} and class II being radiatively pumped \citep{Menten1991, Cragg1992}. The class I CH$_{3}$OH masers are typically associated with interstellar shocks. Especially for protostellar systems, they are related to shocked material in outflows and can emit even at distant locations from protostars \citep{Voronkov2006, Cyganowski2009, Leurini2016, Baek2023}. In contrast, the class II CH$_{3}$OH masers arise in the vicinity of massive protostars due to IR photons from warm dust \citep{Cragg1992, Slysh1994, Sobolev1997, Cragg2002, Baek2023, Hunter2026}.\par

Since the first discovery of CH$_3$OH maser emission towards Orion-KL \citep{Barrett1971}, most studies have been done on low frequency transitions in the frequency range from a few to several tens of gigahertz (e.g. \citealt{Wilson1984, Menten1991, Pestalozzi2006, Voronkov2011, Breen2019}). Over the past two decades, however, studies of CH$_3$OH masers have expanded into the millimetre/submillimetre regime, revealing maser activity at frequencies of a few hundred gigahertz (e.g. \citealt{Ellingsen2012, Zinchenko2017, Chen2019, Humire2022, Wakelam2022, Baek2023}).\par

In particular, for class~I CH$_3$OH masers, single dish and interferometric studies have reported maser emission in several millimetre/submillimetre transitions, including those at 181.3, 218.4, 229.8, 250.5, 266.8, 278.3, and 326.9~GHz \citep{Cyganowski2011,Cyganowski2012,Cyganowski2017,Cyganowski2018,Voronkov2012,Hunter2014,Yanagida2014,Cunningham2016,Chen2019,Kim2020,Chahine2022_maser,Hirota2022,Baek2023,Williams2023,Humire2024}. Collectively, these studies indicate that millimetre/submillimetre class I CH$_3$OH masers arise in shocked gas associated with protostellar outflows, bow shocks, cavity walls, or outflow-ambient interaction regions.\par

Most spatially resolved studies of class I CH$_3$OH masers reported so far have focused on high-mass star-forming regions. In contrast, studies of low-mass sources have been dominated by single dish surveys \citep{Kalenskii2006, Kalenskii2010, Kalenskii2012, Kalenskii2013, Gan2013, Kang2013, Lyo2014, Humire2024}, while interferometric observations that spatially resolve the masing gas in low-mass protostars have remained limited \citep{Kalenskii2017, Bouvier2025}. Interferometric observations of class I CH$_3$OH masers towards a uniformly observed sample of low-mass protostars therefore provide a valuable opportunity to investigate their occurrence, spatial origin, and physical properties in this mass regime.\par

The model-predicted class I CH$_{3}$OH maser $12_{0,12}-11_{1,11}$A transition ($E_{\rm u}$ = 180.9~K, $\nu$ = 302.913 GHz, log$_{10}(A_{\rm ul}(\rm s^{-1}))=-3.82$) has not been identified as a maser in the interstellar medium. However, a calculation of energy level populations predicted it to be a maser line, based on models of non-dissociative magnetohydrodynamic C-type shock waves propagating through the neutral gas component \citep{Nesterenok2021}.\par

In this paper, we report the first detection of the 302.9~GHz CH$_{3}$OH maser, identified serendipitously in the spectral survey data of the Complex Organic Molecules in Protostars with ALMA Spectral Surveys (COMPASS). COMPASS is a Large Program carried out with the Atacama Large Millimeter/submillimeter Array (ALMA) and designed to obtain a comprehensive chemical inventory of 11 embedded low-mass young stellar objects harbouring hot corinos. An overarching goal of COMPASS is to understand how protostellar environments and evolutionary stages regulate the formation and evolution of COMs. Because COMPASS data provide uniform interferometric spectral coverage across multiple CH$_3$OH transitions in embedded low-mass protostars, these data allow us to assess the occurrence of this newly identified submillimetre class I CH$_3$OH maser transition and localise the masing gas within these systems.\par

COMPASS data also enable us to investigate the physical and chemical conditions of the shocked gas in which class I CH$_3$OH masers arise. Because class I CH$_3$OH masers are collisionally pumped in shocks, they provide sensitive tracers of shocked gas in protostellar systems. Regions with detected maser emission can therefore be used as laboratories for shock chemistry. By examining which chemical processes operate there and which COMs are produced, we can directly probe the chemistry driven by protostellar shocks. In particular, by analysing the rich set of COM transitions present in our spectral survey, we identify which COMs are preferentially associated with the CH$_3$OH masing gas and characterise their spatial distribution. Conversely, the associated shock-driven chemical signatures can provide strong evidence of the shock origin of the detected class I CH$_3$OH maser emission.\par

This paper is organised as follows. In Sect.~\ref{sec:observation}, we describe the observations. In Sect.~\ref{sec:Results}, we present a comparative analysis of the emission morphology of the 302.9~GHz class I CH$_3$OH maser candidate and those of other CH$_3$OH and H$_2$CO thermal transitions. In Sect.~\ref{sec:rotation diagram}, we perform a rotation diagram analysis to identify the sources that exhibit non-local thermodynamic equilibrium (non-LTE) excitation in the 302.9~GHz emission and determine where the maser emission is predominantly found within the low-mass protostellar systems. In Sect.~\ref{sec:discussion}, we compare the properties of sources with and without maser detections to investigate the evolutionary and environmental conditions favourable for the 302.9~GHz CH$_3$OH maser. We also show that gaseous acetaldehyde (CH$_3$CHO) is enriched along the masing components around HOPS 108, likely due to protostellar shocks that also excite the 302.9~GHz CH$_3$OH maser. Finally, we summarise our results in Sect. \ref{sec:summary}.

\section{Observations} \label{sec:observation}

    \begin{table*}[htbp]
    \centering
    \caption{Basic information about the COMPASS targets and image cubes.}
    \label{tb:observation}
    \begin{tabular}{llccccc}
        \hline
        Target & Abbr. name\textsuperscript{(a)} & Distance\textsuperscript{(a)} &$T_{\rm bol}$\textsuperscript{(a)} & $v_{\rm lsr}$\textsuperscript{(a)} & Line HPBW\textsuperscript{(b)} & Line HPBW \\ 
               &                               &    (pc)                     & (K)                & (km s$^{-1}$)              & min$-$max ($\arcsec$)        & min$-$max (au)\\
        \hline\hline
        BHR71-IRS1 & BHR71IRS1 & 176 & 49 & -4.6 & $0.36-0.44$ & $63-77$ \\
        HOPS 108 & HOPS108 & 393 & 53 & 12.2 & $0.28-0.41$ & $110-161$ \\
        HOPS 373-B & HOPS373 & 417 & 36 & 9.9 & $0.28-0.42$ & $117-175$ \\
        NGC 1333-IRAS4A & IRAS4A & 290 & 27 & 6.8 & $0.26-0.48$ & $75-139$ \\
        Serpens emb 11 W & Emb11W & 436 & 49 & 8.5 & $0.26-0.44$ & $113-192$ \\
        Serpens SMM1a & SMM1a & 436 & 38 & 7.5 & $0.25-0.43$ & $109-187$ \\
        Serpens South-MM18a & MM18a & 441 & 37 & 7.9 & $0.24-0.42$ & $106-185$ \\
        B335 IRS & B335 & 165 & 43 & 6.9 & $0.23-0.49$ & $38-81$ \\
        L1551-IRS5 & L1551IRS5 & 145 & 164 & 9.7 & $0.13-0.16$ & $19-23$ \\
        NGC 1333-SVS13a & SVS13a & 290 & 199 & 7.7 & $0.14-0.18$ & $41-52$ \\
        V883-Ori & V883Ori & 388 & 664 & 4.1 & $0.28-0.41$ & $109-159$ \\
        \hline
        \end{tabular}
    
        \vspace{0.5em}
        \raggedright
        \textbf{Notes.} The maximum recoverable scale is about 15 times larger than the angular resolution.
        \textsuperscript{(a)}Abbreviated name, distance, $T_{\rm bol}$, and $v_{\rm lsr}$ value of the targets are adopted from \citet{jorgensen25compass}.
        \textsuperscript{(b)}Half power beam width (HPBW) values of the synthesised beam are taken from \citet{plunkett25compass}.
    \end{table*}

The observations were conducted as part of the COMPASS ALMA Large Program (2022.1.00316.L; \citealt{jorgensen25compass}),\footnote{https://erda.dk/vgrid/COMPASS/} an unbiased spectral survey of 11 low-mass protostars spanning diverse environments from dense, clustered star-forming regions to isolated cores. The frequency coverage of $279.0-311.7$ GHz in ALMA's Band 7 was chosen as an optimal compromise between molecular line density and atmospheric transmission, enabling detection of key complex organics and rare isotopologues. Most sources were observed at $0\farcs33-0\farcs40$ angular resolution, sufficient to isolate the warm inner envelope emission from the larger-scale outflows. Two tight binaries (SVS13a, L1551IRS5) were targeted at $\sim0\farcs15$ resolution to separate the components. The sources are located at distances ranging from 145 to 441 pc \citep{jorgensen25compass}, and, given the synthesised beam of the observations, the corresponding resolved physical scales range from 19 to 192 au. The velocity resolution is $0.47-0.52$~km s$^{-1}$ (488.281~kHz per channel), and the sensitivity is $2.15-2.69$ mJy beam$^{-1}$.\par

The observations were carried out between January 2023 and August 2024. Basic information about the COMPASS targets and observations is listed in Table \ref{tb:observation}. For detailed descriptions of the observations, data reduction procedures, and general information on the targeted sources, we refer the reader to the COMPASS overview paper \citep{jorgensen25compass} and the dedicated data reduction paper \citep{plunkett25compass}. We also refer to \citet{maret2025} for the COMPASS data reduction pipeline.\par

To better illustrate the emission morphology, the maps shown in all the figures in this paper are presented without primary-beam correction. In contrast, the brightness temperatures annotated in the figures and used throughout the analysis are derived from primary-beam-corrected data. Also, these brightness temperatures represent beam-averaged quantities. We did not apply any beam-dilution correction for the emission size.

\section{Morphological comparison of line emission} \label{sec:Results}
\subsection{302.9~GHz CH$_3$OH maser candidate detection} \label{sec:maser map}

    \begin{figure*}[t]
        \centering
        \includegraphics[trim=0cm 0cm 0cm 0cm, clip=true,width=0.85\textwidth]{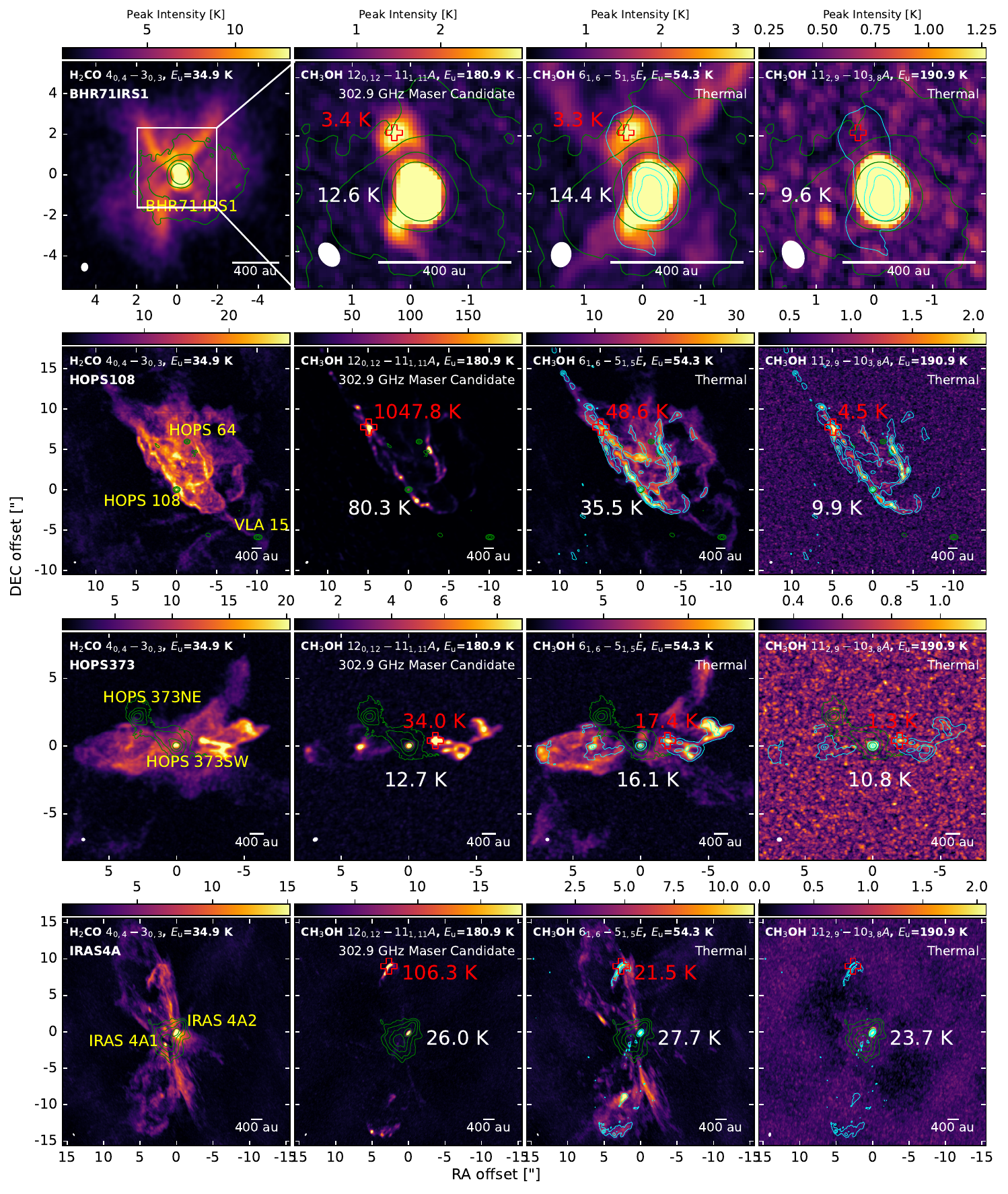}
        \caption{Peak $T_{\rm b}$ maps of an H$_{2}$CO line (log$_{10}(A_{\rm ul}(\rm s^{-1}))=-3.16$, 290.6~GHz) and three CH$_{3}$OH lines (302.9~GHz, log$_{10}(A_{\rm ul}(\rm s^{-1}))=-3.82$; 290.1~GHz, log$_{10}(A_{\rm ul}(\rm s^{-1}))=-3.99$; and 281.0~GHz, log$_{10}(A_{\rm ul}(\rm s^{-1}))=-4.45$) towards the COMPASS sources. The second column shows the 302.9~GHz CH$_3$OH maser candidate. For sources where the maser candidate emission is confined to a smaller region than the H$_2$CO emission, a zoomed-in view within the H$_2$CO field of view is presented. In the third and fourth columns, cyan contours show the peak $T_{\rm b}$ distribution of the maser candidate at contour levels of 5, 20, 35$\sigma$ for comparison. Green contours trace the 1 mm continuum emission at contour levels of 10, 30, 90, and 270$\sigma$. The red cross marks the brightest pixel of the maser candidate outside the central hot corino, as identified in the second column; the same position is also indicated in the third and fourth columns. Red numbers indicate the $T_{\rm b}$ of each CH$_{3}$OH transition at the red cross position when detected above 4$\sigma$. White numbers indicate the $T_{\rm b}$ of each transition measured at the reference positions of central hot corinos defined by \citet{plunkett25compass}, which are near the coordinate origin. The displayed images are not primary-beam corrected, but the red and white $T_{\rm b}$ values are reported after primary-beam correction. The synthesised beam is shown as a white ellipse in the bottom-left corner of each panel.}
        \label{fig:h2co_methanol mom8 map}
    \end{figure*}

    \begin{figure*}[t]
        \centering
        \includegraphics[trim=0cm 0cm 0cm 0cm, clip=true,width=0.85\textwidth]{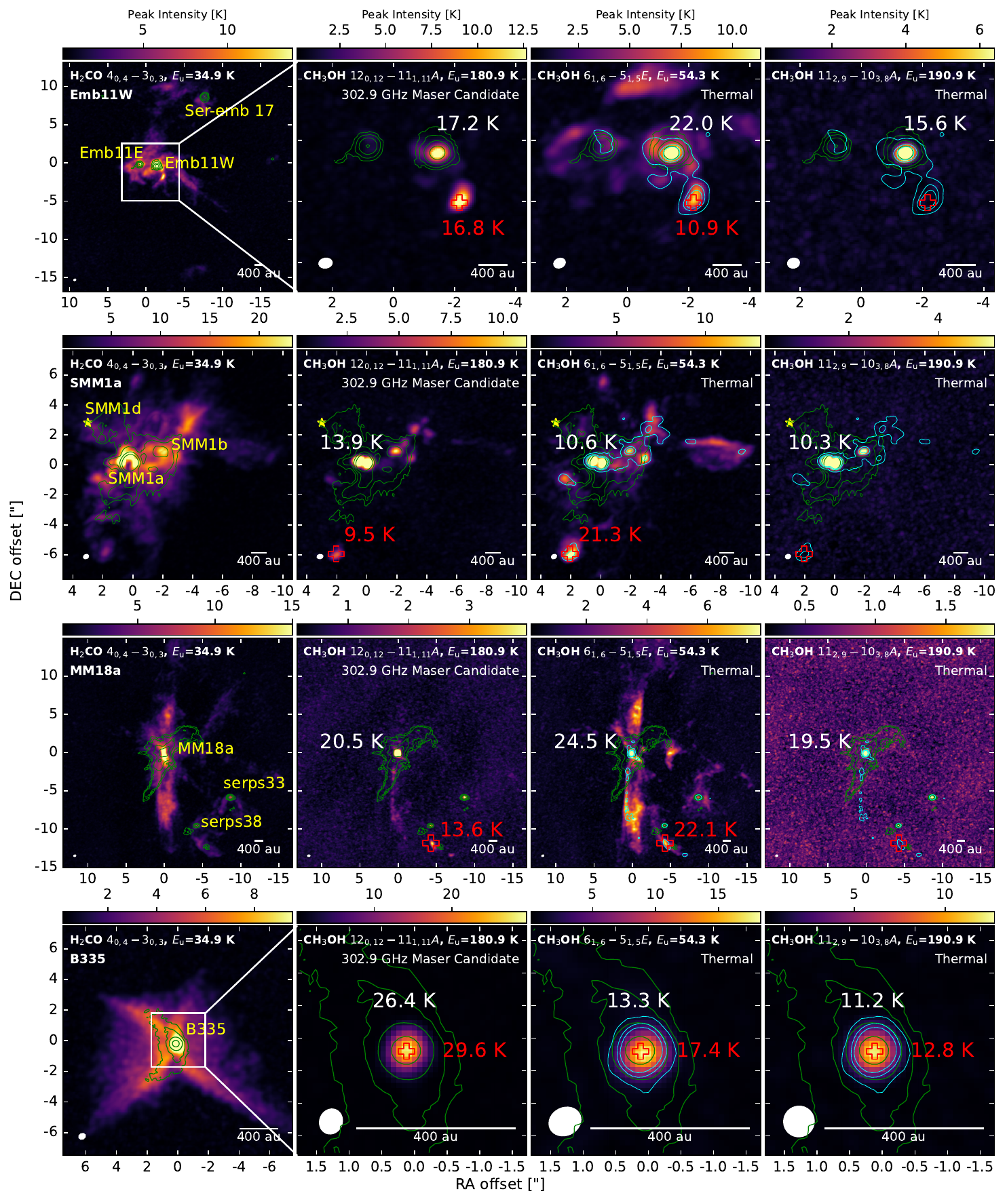}
        \addtocounter{figure}{-1}
        \caption{Continued (panels for Emb11W, SMM1a, MM18a, and B335). SMM1d, one of the outflow-driving sources in the SMM1 field \citep{Hull2017}, is marginally visible at these contour levels. Its position is marked with a yellow star.}
    \end{figure*}

    \begin{figure*}[t]
        \centering
        \includegraphics[trim=0cm 0cm 0cm 0cm, clip=true,width=0.85\textwidth]{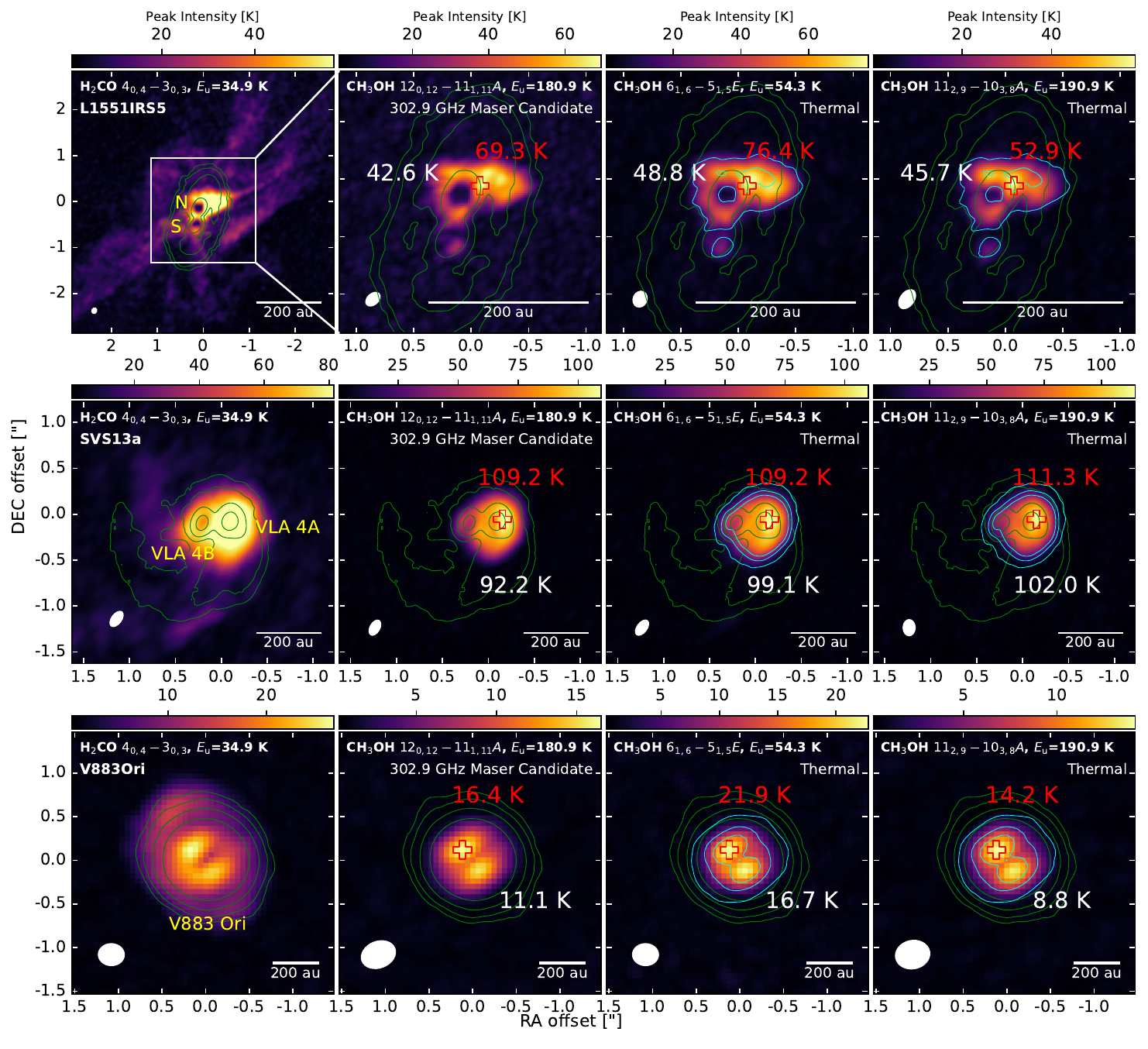}
        \addtocounter{figure}{-1}
        \caption{Continued (panels for L1551IRS5, SVS13a, and V883Ori).}
    \end{figure*}

Figure \ref{fig:h2co_methanol mom8 map} presents peak brightness temperature maps of one H$_2$CO transition and three CH$_3$OH transitions. The H$_2$CO $4_{0,4}-3_{0,3}$ transition ($E_{\rm u}=34.9$~K) is included because H$_2$CO is a well-known tracer of protostellar envelopes and outflows \citep{Tychoniec2021}. Its emission map provides an overview of the extended molecular gas structures around each source and serves as a spatial reference for interpreting the CH$_3$OH emission. The three CH$_3$OH panels show the 302.9~GHz transition 12$_{0,12}-11_{1,11}$A, identified as the maser candidate transition in this study, together with two comparison thermal lines: one with a lower upper-state energy ($E_{\rm u}=54.3$~K) and with a similar upper-state energy ($E_{\rm u}=190.9$~K). These comparison maps allowed us to assess how the morphology of the 302.9~GHz emission relates to that of the thermal emission.\par

In general, the H$_2$CO emission consists of compact hot corino components at 1 mm continuum peaks and more extended structures, such as outflow cavity walls or envelopes. The spatial distribution of CH$_3$OH varies across the three transitions and among the 11 COMPASS sources. Throughout this and the following subsection, we refer to the three CH$_3$OH transitions by their upper-state energies $E_{\rm u}$ for brevity.\par

Based on the morphological comparisons, the 11 sources can be divided into two groups: in Group 1, the three CH$_3$OH emission maps show distinct morphologies, whereas in Group 2 they exhibit similar morphologies. Group 1 consists of BHR71IRS1, HOPS108, HOPS373, IRAS4A, Emb11W, SMM1a, and MM18a, while Group 2 consists of B335, L1551IRS5, SVS13a, and V883Ori.\par

In the seven Group 1 sources, the $E_{\rm u}=54.3$~K transition shows extended molecular structures that closely resemble those seen in H$_2$CO. By contrast, the CH$_3$OH maser candidate $E_{\rm u}=180.9$~K line appears as several knots at outflow lobe tips (bow shocks) and along cavity walls, sampling only a subset of the extended emission of the 54.3~K line. The $E_{\rm u}=190.9$~K transition is generally confined to the hot corino regions near the continuum peaks. In contrast, the four Group 2 sources show no compact knots offset from the central sources: instead, the three CH$_3$OH transitions exhibit similar emission morphologies and are primarily confined to the hot corino region, with little or no extension along the outflow cavities.\par

We also compared the brightness temperatures of the transitions focusing on two pixels, indicated by the red and white numbers in Fig.~\ref{fig:h2co_methanol mom8 map} (see also Table~\ref{tb:brightness temperature}). The red numbers indicate the brightness temperature measured at the brightest position in the 180.9~K map (marked by a red cross); to properly explore the extended emissions, we excluded the central hot corino region when searching for the brightest spot for the Group 1 sources. Among the Group 1 sources, the brightness temperature of the 180.9~K transition at the red cross varies relative to the 54.3~K line: it is brighter in HOPS108, HOPS373, IRAS4A, and Emb11W; fainter in SMM1a and MM18a; and comparable in BHR71IRS1. In all these cases, however, the 190.9~K transition is either undetected or very weak at this position. In contrast, the four sources in Group 2 exhibit comparable brightness temperatures in all three transitions, except for B335. In B335, the 54.3 and 190.9 K transitions exhibit comparable intensities, whereas the 180.9 K transition is more than twice as bright as either of them. This difference is discussed further in Sect. \ref{sec:individual sources}.\par

The white numbers denote the line brightness temperatures measured at the reference positions defined by \citet{plunkett25compass}, which correspond to the continuum peak of the main hot corino in each COMPASS source. For L1551IRS5 only, a position offset by $0\farcs15$ to the north of L1551IRS5~N is adopted, due to the high dust optical depth at the continuum peak \citep{plunkett25compass}. Most sources show relatively similar line brightness temperatures among the three transitions at these reference positions. An exception is HOPS108, where the 180.9~K transition reaches a brightness temperature over eight times that of the 190.9~K line. This is discussed in more detail in the context of the rotation diagram analysis (Sect. \ref{sec:ratio map}). The positions and brightness temperatures used in this analysis are listed in Table~\ref{tb:brightness temperature}.\par

    \begin{table*}[htbp]
    \centering
    \caption{Reference positions used in Fig. \ref{fig:h2co_methanol mom8 map} and the brightness temperatures ($T_{\rm b}$) measured at those positions.}
    \label{tb:brightness temperature}
    \scriptsize
    \begin{tabular}{lcccccccc}
        \hline
        Target & Maser candidate peak\textsuperscript{(a)} & \multicolumn{3}{c}{$T_{\rm b}$ (K)\textsuperscript{(b)}} & Continuum peak\textsuperscript{(c)} & \multicolumn{3}{c}{$T_{\rm b}$ (K)\textsuperscript{(d)}} \\ 
               &  RA DEC (ICRS) & 180.9 K & 54.3 K & 190.9 K & RA DEC (ICRS) & 180.9 K & 54.3 K & 190.9 K  \\
        \hline\hline
        BHR71IRS1 & 12:01:36.547(9) -65:08:48.36(6) & 3.4 & 3.3& $<0.69$\textsuperscript{(e)} & 12:01:36.4750(13) -65:08:49.361(13) & 12.6& 14.4& 9.6\\
        HOPS108 & 05:35:27.408(3) -05:09:52.42(5) & 1047.8 & 48.6 & 4.5 & 05:35:27.0847(7) -05:10:00.08(1) & 80.3& 35.5& 9.9\\
        HOPS373 & 05:46:30.767(3) -00:02:34.80(5) & 34.0 & 17.4 & 1.3 & 05:46:30.9050(9) -00:02:35.18(1) & 12.7& 16.1& 10.8\\
        IRAS4A & 03:29:10.654(2) +31:13:41.03(3) & 106.3 & 21.5 & $<1.3$ & 03:29:10.4380(9) +31:13:32.03(3) & 26.0& 27.7& 23.7\\
        Emb11W & 18:29:06.575(3) +00:30:32.30(5) & 16.8 & 10.9 & $<0.96$ & 18:29:06.6247(5) +00:30:33.89(1) & 17.2& 22.0& 15.6\\
        SMM1a & 18:29:49.925(3) +01:15:14.20(5) & 9.5 & 21.3 & $<1.1$ & 18:29:49.8037(5) +01:15:20.34(1) & 13.9& 10.6& 10.3\\
        MM18a & 18:30:03.825(3) -02:03:14.45(5) & 13.6 & 22.1 & $<2.9$ & 18:30:04.1227(7) -02:03:02.64(1) & 20.5& 24.5& 19.5\\
        B335 & 19:37:00.902(4) +07:34:09.41(6) & 29.6 & 17.4 & 12.8 & 19:37:00.9027(13) +07:34:09.41(3) & 26.4& 13.3& 11.2\\
        L1551IRS5 & 04:31:34.152(2) +18:08:04.68(2) & 69.3 & 76.4 & 52.9 & 04:31:34.1656(5) +18:08:04.59(1) & 42.6& 48.8& 45.7\\
        SVS13a & 03:29:03.742(1) +31:16:03.73(2) & 109.2 & 109.2 & 111.3  & 03:29:03.7491(7) +31:16:03.718(7) & 92.2& 99.1& 102.0\\
        V883Ori & 05:38:18.106(3) -07:02:25.90(5) & 16.4 & 21.9 & 14.2 & 05:38:18.0998(6) -07:02:25.986(11) & 11.1& 16.7& 8.8\\
        \hline
        \end{tabular}
    
        \vspace{0.5em}
        \raggedright
        \textbf{Notes.} The $T_{\rm b}$ values listed to the right of each reference position correspond, from left to right, to the three CH$_3$OH transitions shown in Fig.~\ref{fig:h2co_methanol mom8 map}, with $E_{\rm u}=$180.9 K, 54.3 K, and 190.9 K, respectively. The reported $T_{\rm b}$ values were measured from primary-beam-corrected images.\\
        \textsuperscript{(a)} Because each maser candidate peak position was taken directly from the brightest image pixel, its positional uncertainties have been adopted as the corresponding pixel sizes. The numbers in parentheses indicate the positional uncertainties in the last quoted digits. \textsuperscript{(b)} These values are shown in red in Fig.~\ref{fig:h2co_methanol mom8 map}.\\ \textsuperscript{(c)} The continuum peak positions are the median coordinates obtained from two-dimensional Gaussian fits to the continuum emission in the 36 spectral windows \citep{plunkett25compass}. The uncertainties are the rms dispersions among the fitted positions. \textsuperscript{(d)} These values are shown in white in Fig.~\ref{fig:h2co_methanol mom8 map}.\\
        \textsuperscript{(e)} Upper limits for undetected lines indicate the 4$\sigma$ noise level.
    \end{table*}

\subsection{Individual sources}\label{sec:individual sources}
In this subsection, we present source-by-source descriptions of the H$_2$CO and CH$_3$OH peak brightness temperature maps (Fig. \ref{fig:h2co_methanol mom8 map}). For each source, we first describe how the extended H$_2$CO emission relates to the outflow structures identified in previous CO studies, and then we compare the emission morphologies of the three CH$_3$OH transitions. For general source properties, see the COMPASS overview paper \citep{jorgensen25compass}.

\paragraph{BHR71-IRS1:} 
BHR71IRS1 is one of the only two isolated cores in the COMPASS sample. The H$_2$CO map exhibits relatively simple geometry, with extended emission that is consistent with the outflow structure identified in previous CO studies \citep{Zapata2018, Tobin2019_2}. In the zoomed-in view, the CH$_3$OH $E_{\rm u} = 180.9$~K line map reveals extensions immediately north and south of the central hot corino, while no emission is detected in more distant regions such as the outflow cavity wall of BHR71IRS1. The overall large-scale outflow is oriented north-south, with the northern lobe redshifted and the southern lobe blueshifted. In contrast, the northern CH$_3$OH emission components are blueshifted (see Fig. \ref{fig:spectra1} in Sect. \ref{sec:Rotation diagram fitting result} for the spectra), indicating that they do not follow the velocity pattern of the overall large-scale outflow \citep{Yang2020, yun25compass}. The 54.3~K CH$_3$OH transition also traces these components, as well as the outflow cavity walls that are seen in the H$_2$CO map. The emission of the 190.9~K transition is confined to the hot corino region.\par

\paragraph{HOPS 108:} 
The H$_2$CO structure reflects the outflow driven by the intermediate-mass protostar HOPS370 with a bolometric luminosity of 360.9 $L_{\odot}$ \citep{Adams2012, Furlan2014, Furlan2016, Tobin2019, Chahine2022_maser, Sato2023}. HOPS370 is located $\sim$27$\arcsec$ northeast from HOPS108, which is outside the observed field of the COMPASS programme (see Fig. \ref{fig:ch3cho large map} in Sect. \ref{sec:CH3CHO} for a larger field of view). There are only tentative detections of outflow powered by HOPS108 \citep{Tobin2019, Sato2023}.\par

The CH$_3$OH 180.9~K map reveals several compact, bright (brightest of all COMPASS sources), spot-like features distributed along the outflow lobe of HOPS370 oriented northeast–southwest. The brightest spot with a brightness temperature of 1047.8~K lies approximately $1\farcs8$ from the VLA 12N jet knot detected in the 3 and 5 cm continua \citep{Osorio2017}, and it coincides with the P3 maser spot identified in the class I CH$_3$OH maser $4_{-2,3}-3_{-1,2}$E ($E_{\rm up}=45.5$~K, 218.440~GHz) transition by \citet{Chahine2022_maser}.\par

The $E_{\rm u} = 54.3$~K transition exhibits more extended CH$_3$OH distribution that encompasses the emission seen in the 180.9~K map, indicated with cyan contours, but it is generally fainter and shows a different brightness distribution. At the red cross, for example, its brightness temperature is approximately 20 times lower than that of the 180.9~K line. The 190.9~K transition, in contrast to BHR71IRS1, shows weak, but spatially extended emission in HOPS108. However, its brightness temperature remains relatively low, being only a few kelvin.

\paragraph{HOPS 373-B:} 
The dominant H$_2$CO structures are associated with outflows from HOPS373SW rather than the northeastern binary component HOPS373NE \citep{Tobin2020, Yoon2022, SHLee2024}. The CH$_3$OH 180.9~K map shows brighter knot structures along the western outflows. A similar structure is seen in Gemini North/GNIRS K-band observations, which only find H$_2$ molecular emission towards the western side \citep{Yoon2022}. Also, \citet{SHLee2024} studied the outflow features of HOPS373SW using CH$_3$OH $7_{\rm k}-6_{\rm k}$ thermal transitions at $338.34-338.73$ GHz. They identified three strong CH$_{3}$OH redshifted emission components (CR1, CR2, CR3) and three blueshifted ones (CB1, CB2, CB3) along the wings of the bow shocks of two small-scale outflows (Outflow1 and Outflow2, see Fig. 2 of \citealt{SHLee2024}). Except for CR1, which has the lowest CH$_3$OH column density \citep{SHLee2024} among the six bright CH$_3$OH components, all other regions appear as bright spots in the 180.9~K peak brightness temperature map, where the position of the red cross coincides with that of CB1.\par

The 54.3~K map also traces structures extending towards the northwest, southwest, and southeast, which are not visible in the 180.9~K emission, but are apparent in the H$_2$CO map. The 190.9~K map shows bright emission only towards the hot corino of HOPS373SW, without any extended structure. The absence of CH$_3$OH emission from HOPS373NE is consistent with the results of \citet{JELee2023}.\par

\paragraph{NGC 1333-IRAS4A:}
In IRAS4A, multiple lobes can be seen in the H$_2$CO map, corresponding to outflows from both Class 0 components IRAS4A1 and IRAS4A2 \citep{Santangelo2015, Chuang2021}. IRAS4A is one of the few low-mass protostars where a class I 44~GHz CH$_3$OH maser has been detected: NGC 1333–IRAS 4A, NGC 1333–IRAS 2A, HH 25 MMS, and L1157 \citep{Kalenskii2010}. \citet{Lyo2014} reported the detection of the class I 132.9~GHz CH$_3$OH maser towards IRAS4A with the 21 m Korean VLBI Network (KVN) in the single-dish mode at an angular resolution of 23$\arcsec$. In addition, using the Expanded Very Large Array (EVLA) at 0.2$\arcsec$ resolution, \citet{Kalenskii2017} found 44~GHz CH$_3$OH maser spots south of IRAS4A. These correspond to the southern spots seen in the CH$_3$OH $E_{\rm u}=180.9$~K panel of Fig. \ref{fig:h2co_methanol mom8 map}.\par

The 54.3~K CH$_3$OH emission map, similar to that of H$_2$CO, closely follows the morphology of the CO outflows reported in \citet{Chuang2021}, tracing the cavity structures originating from IRAS4A1 and IRAS4A2. In contrast, the 180.9~K emission is concentrated at the tips of these outflow lobes, delineating arc-like structures that correspond to bow shocks. The 190.9~K transition exhibits emission only towards the hot corino associated with IRAS4A2, similar to other COMs tracing the hot corino \citep{Lopez2017}.\par

\paragraph{Serpens emb 11 W:}
Emb11W exhibits complex H$_2$CO morphologies because there are multiple overlapping outflows. The main outflow-driving sources are Emb11W, Emb11E, and Ser-emb 17 \citep{Djupvik2016, Mart2021}.\par

The CH$_3$OH 54.3~K map likewise shows a complex emission morphology consistent with these multiple outflows, as seen in the H$_2$CO. In contrast to most other fields in our sample, the extended emission of the $E_{\rm u}=180.9$~K transition is confined to the immediate vicinity of the Emb11E/W binary, with no significant features along the more distant outflow structures, resembling the pattern seen in BHR71IRS1. The 190.9~K transition is detected only towards Emb11W within the Emb11E/W binary, consistent with \citet{Mart2021}.\par

\paragraph{Serpens SMM1a:}
Similar to Emb11W, the H$_2$CO map of SMM1a exhibits a complex morphology because the SMM1 field contains overlapping outflows driven by SMM1a, SMM1b, and SMM1d \citep{Hull2017}. The CH$_3$OH $E_{\rm u}=180.9$ and 54.3~K maps likewise show complex morphologies shaped by these multiple outflows, revealing several bow shock structures. As an example, the westernmost isolated bow shock structure, visible in the three left panels at RA offsets between about -10$^{\arcsec}$ and -6$^{\arcsec}$ and Dec offsets between about 0$^{\arcsec}$ and 2$^{\arcsec}$, corresponds well to the SiO emission associated with the SMM1d-driven outflow \citep{Hull2017}. The other extended CH$_3$OH structures are likely linked to SMM1a, as the H$_2$CO morphology observed by \citet{Tychoniec2019} indicates that, aside from the SMM1d-driven western bow shocks, the H$_2$CO emission mostly traces the SMM1a outflow.\par

However, the CH$_3$OH $E_{\rm u}=180.9$ and 54.3~K maps are not identical in their spatial distribution: compared to the 180.9~K emission, the 54.3~K emission is more spatially extended and shows a more prominent western bow shock structure, as well as brighter emission in the southeastern component marked by the red cross. By contrast, the 180.9 K emission is more localised around compact knots. Lastly, emission in the 190.9~K line is observed exclusively at the positions of SMM1a and SMM1b. Recently, \citet{Humire2024} reported a single-dish detection of the collisionally pumped class I CH$_3$OH $10_{1,10}-9_{-0,9}$E (326.9~GHz) maser towards Serpens FIRS (which corresponds to SMM1a) at 19.2$\arcsec$ resolution.\par

\paragraph{Serpens South-MM18a:}
The H$_2$CO emission map of MM18a shows a well-collimated bipolar outflow originating from the central source \citep{Plunkett2015, Podio2021}, while weaker outflow structures associated with nearby protostars are also present in the field. While the CH$_3$OH $E_{\rm u}=54.3$~K map predominantly traces the highly collimated bipolar outflow launched by MM18a as well, the 180.9~K emission is concentrated near the base of the northern lobe and along the eastern side of the southern cavity wall. Excluding the hot-corino region, the brightest 180.9~K feature (marked by the red cross) exhibits an arc-like morphology resembling a bow shock, and it may be associated with the southeast outflow driven by the nearby protostar serps33, which also drives a shell-like outflow lobe to the northwest \citep{Plunkett2015, Ortiz2021}. The 190.9~K transition shows emission only towards the central source MM18a and the southwestern dust continuum sources serps33 and serps38 \citep{Plunkett2018}, with no significant extended component. \citet{Humire2024} also reported the class I CH$_3$OH $10_{1,10}-9_{-0,9}$E (326.9~GHz) maser detection towards CARMA-7 (which corresponds to MM18a) at 19.2$\arcsec$ resolution.\par

\paragraph{B335 IRS:}
B335 is the other isolated core in the COMPASS sample. Its H$_2$CO map shows a relatively simple morphology, with extended emission tracing cavity wall structures that are spatially consistent with the outflow identified in previous CO studies \citep{Bjerkeli2019, Kim2024}. Unlike H$_2$CO, the emission of the three CH$_3$OH transitions in B335 is all concentrated towards the central hot corino region. Thus, the brightest pixel of the 180.9~K line emission (red cross) coincides closely with the continuum peak of the central hot corino, resulting in nearly identical brightness temperature values indicated with red and white numbers in Fig. \ref{fig:h2co_methanol mom8 map}. The spatial extents of the three CH$_3$OH transitions are consistent with an emitting area of the hot corino that expanded during the luminosity burst that peaked in 2018, and which appears to have remained extended even after the source returned to its pre-burst luminosity in 2023 \citep{Evans2023, JELee2025}.\par

A notable feature is that the 180.9~K line brightness temperature towards the hot corino of B335 is approximately twice as bright as the other two CH$_3$OH transitions, unlike the typical case in hot corinos, where the three CH$_3$OH lines have comparable intensities. This may indicate the presence of non-thermal emission, potentially driven by an accretion shock, as proposed by \citet{Okoda2022, Okoda2024}. Further confirmation is needed using observations at higher angular resolution.\par

\paragraph{L1551-IRS5:}
For the three Class I protostars (L1551IRS5, SVS13a, and V883Ori) in the COMPASS sample, the H$_2$CO emission is mainly associated with circumbinary or circumstellar disk structures. In L1551IRS5, bright H$_2$CO emission is seen along the northwestern side of the circumbinary disk \citep{Bianchi2020, Takakuwa2020, Cruz2025}, and it likely traces the cavity walls of the wide-angle bipolar outflow \citep{White2006, Stojimirovic2006, Wu2009}.\par

All three CH$_3$OH transitions trace only the circumbinary disk rather than the outflow cavities that appear in the H$_2$CO map. The spatial distribution of CH$_3$OH is consistent with the ALMA Band~6 CH$_3$OH emission reported in \citet{Cruz2025}. Its morphology resembles that of CS reported by \citet{Takakuwa2020}. The emission is confined to the circumbinary disk and is brighter on the western side of source N. This likely reflects the blueshifted outflow cavity on that side, which permits stronger irradiation from source N to escape along the cavity and heat the western circumbinary disk surface \citep{Takakuwa2020}. A central depression coincident with the continuum peak of source N is consistent with optically thick dust that suppresses line emission \citep{Takakuwa2020}.

\paragraph{NGC 1333-SVS13a:}
The H$_2$CO emission in SVS13a traces the circumbinary disk of the 0.3$\arcsec$-separated binary sources VLA 4A and VLA 4B \citep{Anglada2000, Tobin2016}. It also reveals spiral structures different from those traced by the dust continuum, shown as a 10$\sigma$ level contour in the corresponding panel of Fig. \ref{fig:h2co_methanol mom8 map}.\par

The three CH$_3$OH transitions exhibit comparable brightness temperatures and similar emission morphologies, forming an elliptical structure around the circumbinary system. This morphology is consistent with that of CS, H$^{13}$CN, HC$^{15}$N, and CH$_3$CN reported by previous studies \citep{Diaz-Rodriguez2022, Hsieh2025}. The weaker emission towards the eastern protostellar component, VLA 4B, is likely due to its higher dust optical depth \citep{Diaz-Rodriguez2022}.\par

The spiral structures seen in H$_2$CO, which differ from those traced by the dust continuum, are not detected in the CH$_3$OH maps. In addition, a shock-related 22.2~GHz water maser has been reported towards SVS13a \citep{Szabo2023}, and shock tracers such as [Fe II] and H$_2$ 1–0 S(1) emission have been observed within $0.5\arcsec$ southeast of the source \citep{Hodapp2014}. However, none of these features are traced in the 180.9~K CH$_3$OH transition.\par

\paragraph{V883-Ori:}
The H$_2$CO map shows the emission of the Keplerian-rotating inner disk and a fainter, ring-like outer structure as shown in \citet{JELee2024}. All three CH$_3$OH maps of V883~Ori trace the Keplerian disk within $r\sim 0\farcs3$–$0\farcs4$ (i.e. the inner region seen in the H$_2$CO map). The emitting region coincides with the part of the disk where ices have sublimated - water \citep{Tobin2023} and COM, including CH$_3$OH \citep{vantHoff2018, JELee2019}. The central depression in the emission maps is likely due to optically thick dust within the innermost $\sim0\farcs1$ region, which obscures the molecular line emission \citep{Cieza2016, vantHoff2018, JELee2019, rodriguez2022}.\par

\section{Verification of non-LTE excitation} \label{sec:rotation diagram}

Unlike the well-established CH$_3$OH maser transitions at 6.7~GHz (class~II)
and 44~GHz (class~I), the newly identified 302.9~GHz class~I transition requires confirmation that it is masing. In particular, we must establish that the intensity of the line departs from LTE, as expected for maser action. We therefore performed a rotation diagram analysis \citep{Goldsmith1999} to determine which sources exhibit maser emission.\par

\subsection{Line selection for rotation diagrams}
The COMPASS spectral coverage includes a wide range of CH$_3$OH transitions, spanning $E_{\rm u}$ from 16.9 to over 2,000~K. We excluded transitions with low $E_{\rm u}$ values that are likely to become optically thick as well as very high $E_{\rm u}$ transitions whose emission is likely undetectable in the physical conditions of cold molecular outflows.\par

We also only selected isolated CH$_3$OH lines, identified through a brief line identification process using the full spectra at the two representative pixels introduced in Sect. \ref{sec:maser map}. For extended structures, the spectra are typically dominated by a few simple molecules and only a limited number of complex organic molecules, making the selection of isolated CH$_3$OH lines relatively straightforward.\par

However, in hot corino regions, line confusion becomes more significant due to the presence of molecular lines from many species near CH$_3$OH transitions, as can be seen in unbiased spectral survey analyses of COMs (e.g. \citealt{Belloche2016, jorgensen2020_araa, Jeong2025}). To minimise contamination, we exclude CH$_3$OH lines that are blended with simple molecules and major COMs, such as methyl formate (CH$_3$OCHO), acetaldehyde (CH$_3$CHO), ethanol (C$_2$H$_5$OH), formamide (NH$_2$CHO), methyl cyanide (CH$_3$CN), glycolaldehyde (CH$_2$OHCHO), acetone (CH$_3$COCH$_3$), dimethyl ether (CH$_3$OCH$_3$), and their isotopologues.\par

As a result, we exclude all CH$_3$OH lines with $E_{\rm u}$ below 40~K, and select a total of 10 isolated transitions, which include the 302.9~GHz maser candidate line and the other two lines used for morphological comparisons in Fig. \ref{fig:h2co_methanol mom8 map}. We select commonly isolated CH$_3$OH lines across the COMPASS sources. The CH$_3$OH lines used for the rotation diagram analysis are listed in Table~C.1 (available on Zenodo).\par

    \begin{figure*}[t]
        \centering
        \includegraphics[trim=0cm 0cm 0cm 0cm, clip=true,width=1\textwidth]    {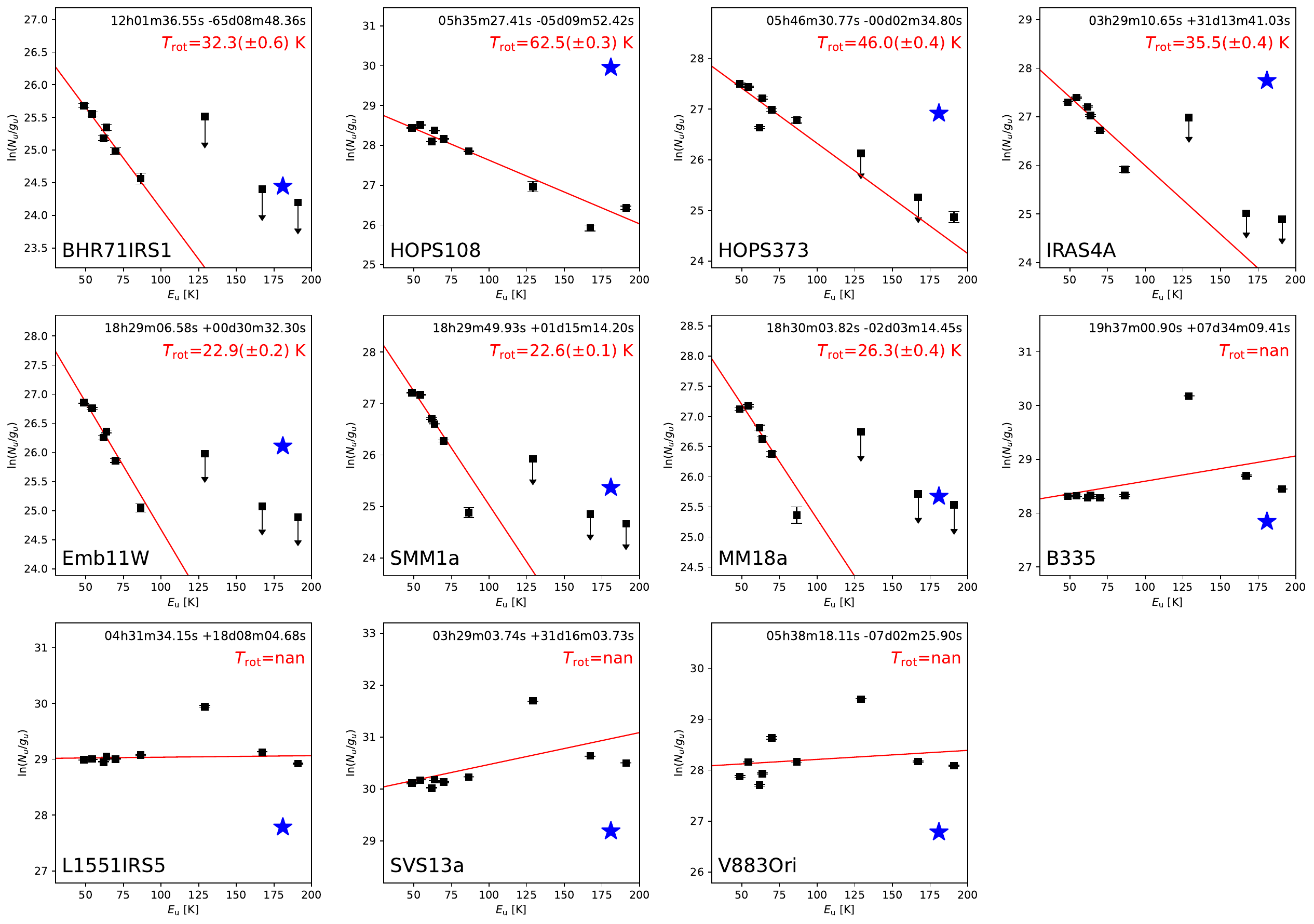}
        \caption{Rotation diagrams for the selected CH$_3$OH transitions at the red cross in Fig.~\ref{fig:h2co_methanol mom8 map}. The red line shows the best-fit model obtained by excluding the 302.9~GHz maser candidate transition (blue point). The ICRS equatorial coordinates of the position, which are listed in Table \ref{tb:brightness temperature}, and the fitted rotation temperature are shown at the top of each panel. Negative temperature solutions due to optically thick lines at low $E_{\rm u}$ are reported as nan (from B335 to V883Ori). Data points with arrows represent upper limits and correspond to transitions whose peak $T_{\rm b}$ values fall below 4$\sigma$.} 
        \label{fig:rotation diagram}
    \end{figure*}

\subsection{Rotation diagram result: Sources with and without maser emission} \label{sec:Rotation diagram fitting result}

    \begin{figure*}[t]
        \centering
        \includegraphics[trim=0cm 0cm 0cm 0cm, clip=true,width=0.495\textwidth] {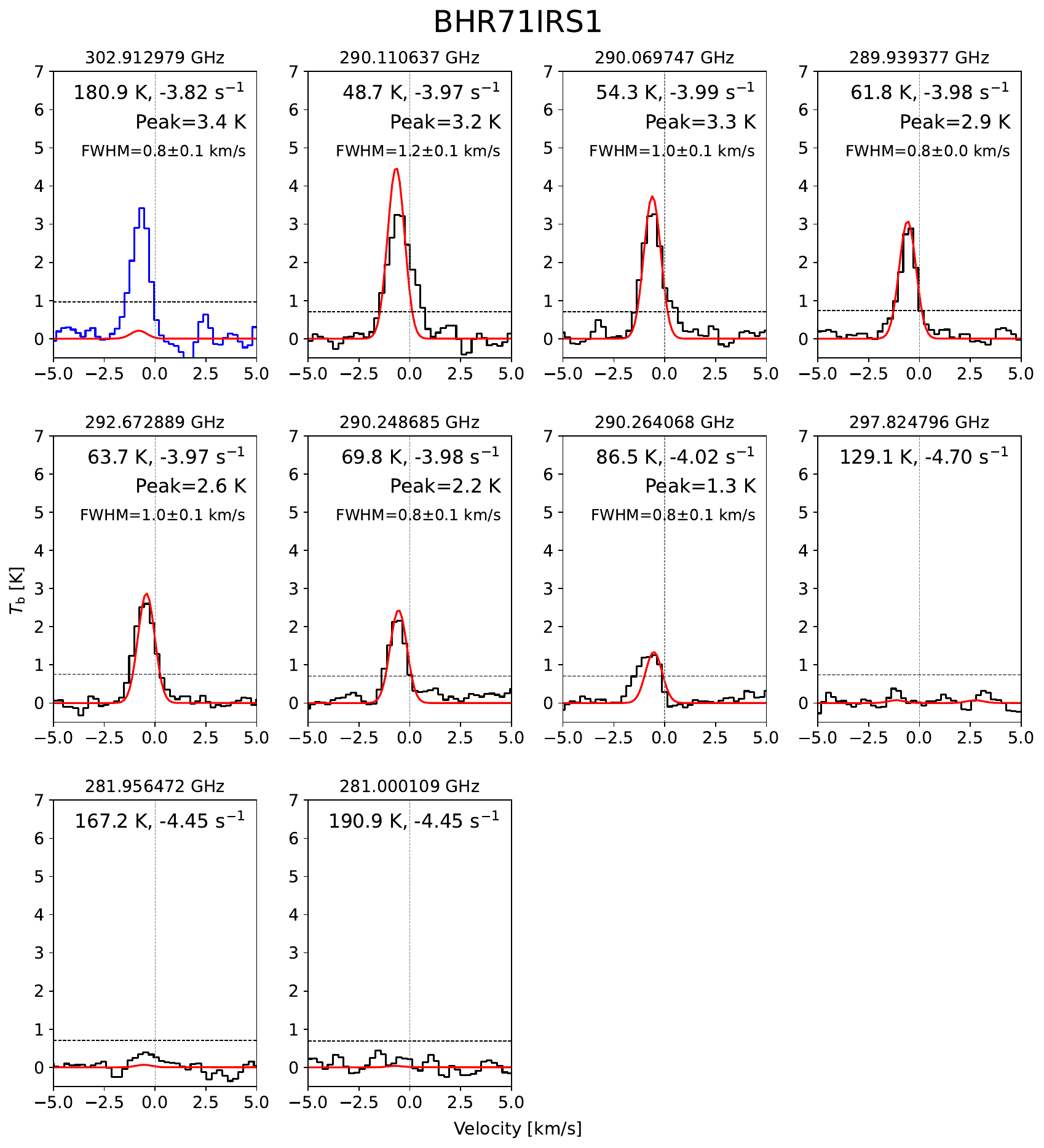}
        \includegraphics[trim=0cm 0cm 0cm 0cm, clip=true,width=0.495\textwidth] {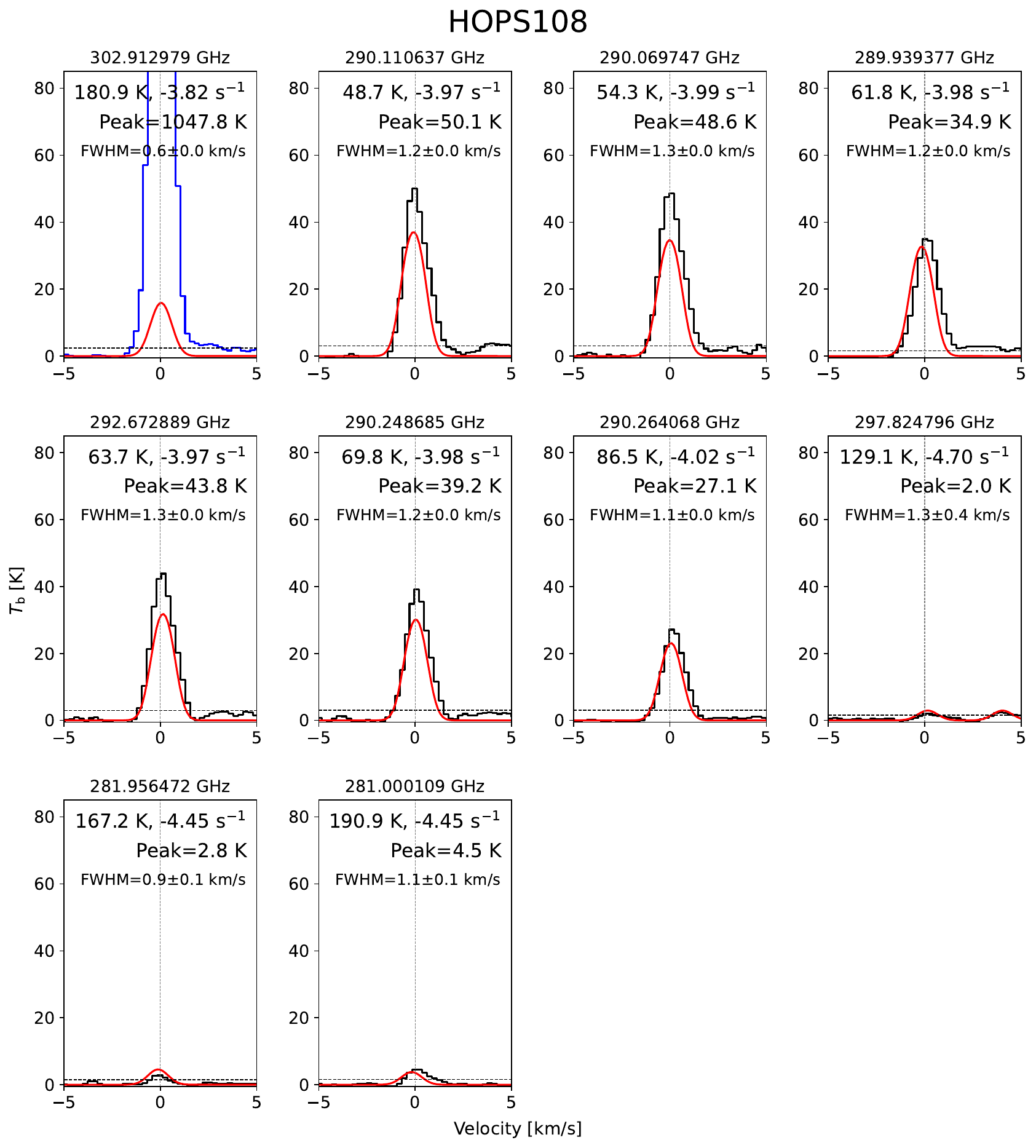}
        \caption{Examples of CH$_{3}$OH spectra (black) extracted from the red cross in Fig.~\ref{fig:h2co_methanol mom8 map} for BHR71IRS1 (left) and HOPS108 (right). The CH$_3$OH lines, except for the 302.9~GHz CH$_3$OH maser candidate in the first panel (blue), are arranged in increasing order of $E_{\rm u}$. The x-axis shows the velocity relative to the systemic velocity, with the grey vertical line marking 0 km s$^{-1}$. The red lines indicate synthetic spectra computed under the assumption of LTE using parameters estimated from the rotation diagram analysis (corresponding panel in Fig. \ref{fig:rotation diagram}). The horizontal dashed line marks the 4$\sigma$ noise level. The two numbers at the top of each panel indicate the $E_{\rm u}$ (in K) and $A_{\rm ul}(\rm s^{-1})$ (in units of log$_{10}$(s$^{-1}$), denoted by s$^{-1}$ in the panels for brevity) values. The peak $T_{\rm b}$ and FWHM of the observed spectra are also provided where the $T_{\rm b}$ value of the transition is higher than 4$\sigma$. The peak $T_{\rm b}$ values are measured directly from the data cubes, whereas the FWHM values are derived from Gaussian fits.}
        \label{fig:spectra1}
    \end{figure*}

Among the selected CH$_3$OH transitions, only those with brightness temperatures exceeding the 4$\sigma$ noise level were fitted with a single Gaussian profile to derive integrated brightness temperatures, which we then used as data points in the rotation diagram. The 302.9~GHz CH$_3$OH maser candidate is excluded when we fit a line to the data points in the rotation diagram. In this subsection, the rotation diagram analysis is carried out using the spectrum extracted at the position of the brightest 302.9~GHz emission (red cross in Fig.~\ref{fig:h2co_methanol mom8 map}).\par

We note that we did not attempt accurate column density estimates here since emission attenuation depends on spatially varying factors such as source size (beam-filling factor) and dust opacity (e.g. \citealt{coutens25compass}), which are not included in this analysis. Our goal was to identify the outliers indicative of masing. Therefore, we neither report the derived column densities nor apply optical depth corrections \citep{Goldsmith1999}, which require accurate column density estimation, in the current rotation diagram analysis.\par

Figure~\ref{fig:rotation diagram} summarises the rotation diagram analysis, confirming maser emission in the seven sources in Group 1 (from BHR71IRS1 to MM18a; Sect. \ref{sec:Results}) and no maser detection in the four sources in Group 2 (from B335 to V883Ori). For Group 1 sources, most data points with $E_{\rm u} > 120$~K are upper limits, except for the 302.9~GHz maser candidate ($E_{\rm u}=180.9$~K), which lies significantly above the thermal fit, deviating strongly from the LTE prediction. The thermal lines themselves follow a single LTE distribution, indicating that LTE holds in these extended structures.\par

On the other hand, the Group 2 sources (from B335 to V883Ori) show two notable effects, without a masing signature. First, no valid rotation temperature can be derived from the rotation diagram. This is likely due to line optical depth effects: lower-$E_{\rm u}$ lines easily become optically thick in dense hot corinos, leading to underestimated $\ln(N_{\rm u}/g_{\rm u})$. This also explains why, in Group 2, the $E_{\rm u}=129.1$ K data point lies above the lower-$E_{\rm u}$ points: the corresponding transition has an $A_{\rm ul}$ about five times smaller and is therefore less susceptible to optical depth effects than the lower-$E_{\rm u}$ transitions. Second, the 302.9~GHz line appears suppressed relative to its neighbouring transitions ($E_{\rm u}=167.2$~K, 190.9~K), also likely due to the line optical depth. The 302.9~GHz line has a $A_{\rm ul}(\rm s^{-1})$ value four to five times higher than those two neighbouring transitions. Thus, if only this transition becomes optically thick, it will appear selectively depleted in the rotation diagram (e.g. \citealt{Jeong2025}).\par

Although the derived column densities for the maser-detected sources are expected to be systematically offset from the true values, they still produce LTE synthetic spectra that reproduce the observed lines. Figure~\ref{fig:spectra1} presents the spectral profiles of CH$_3$OH transitions used in the rotation diagram, along with LTE synthetic spectra generated with the eXtended CASA Line Analysis Software Suite \citep[\texttt{XCLASS};][]{Moller17}, based on rotation temperatures and column densities derived from the rotation diagrams. Two representative cases are shown: HOPS108, where the maser nature of the 302.9~GHz CH$_3$OH line is evident from its exceptionally high brightness temperature, and BHR71IRS1, where a difference in brightness temperature between maser and thermal emission is not apparent. Spectra for the remaining Group 1 sources are shown in Figs.~A.1-A.2 (available on Zenodo); spectra for Group 2 sources are not presented because we cannot derive valid rotation temperatures, as noted above.\par

For all Group 1 sources, the LTE models fail to reproduce the observed brightness temperature of the 302.9~GHz CH$_3$OH maser emission, which explains the discrepancy of the 302.9~GHz data point shown in the rotation diagrams.
For SMM1a and MM18a, the LTE model fails to fully reproduce the thermal CH$_3$OH transitions with $E_{\rm u}=48.7$ and 54.3~K (Fig.~A.2; available on Zenodo). As their observed brightness temperatures are comparable to the derived rotation temperature ($\sim20-30$ K), the lines may not be fully optically thin, and moderate optical depth could account for the discrepancy. Nevertheless, this does not affect our conclusion that the LTE models cannot reproduce the observed 302.9 GHz emission.\par

Kinematically, CH$_3$OH lines have velocities within $\pm$2.5~km~s$^{-1}$ of the $v_{\rm lsr}$ (Table \ref{tb:observation}), suggesting an association with low-velocity shocks. It should be noted, however, that the single-Gaussian fitting does not capture broad wings or multiple velocity components present in some sources, such as HOPS373 (Fig. A.1) and Emb11W (Fig. A.2).\par

Lastly, for the seven maser-detected sources, we performed a 2D Gaussian fit to the brightest compact 302.9~GHz feature (red cross in Fig.~\ref{fig:h2co_methanol mom8 map}) using the CASA \texttt{imfit} task. The deconvolved minor and major axes are 0.13--0.42$^{\arcsec}$ and 0.45--0.99$^{\arcsec}$, respectively. This suggests that the 302.9 GHz maser emission may include marginally resolved maser-emitting structures and/or blends of multiple unresolved maser subcomponents. We therefore do not interpret these sizes as intrinsic sizes of individual maser spots and regard the beam-based brightness temperatures as lower limits.

\subsection{Identifying masing regions using an `$\eta$' map} \label{sec:ratio map}

    \begin{figure*}[t]
        \centering
        \includegraphics[trim=0cm 0cm 0cm 0cm, clip=true,width=1\textwidth]{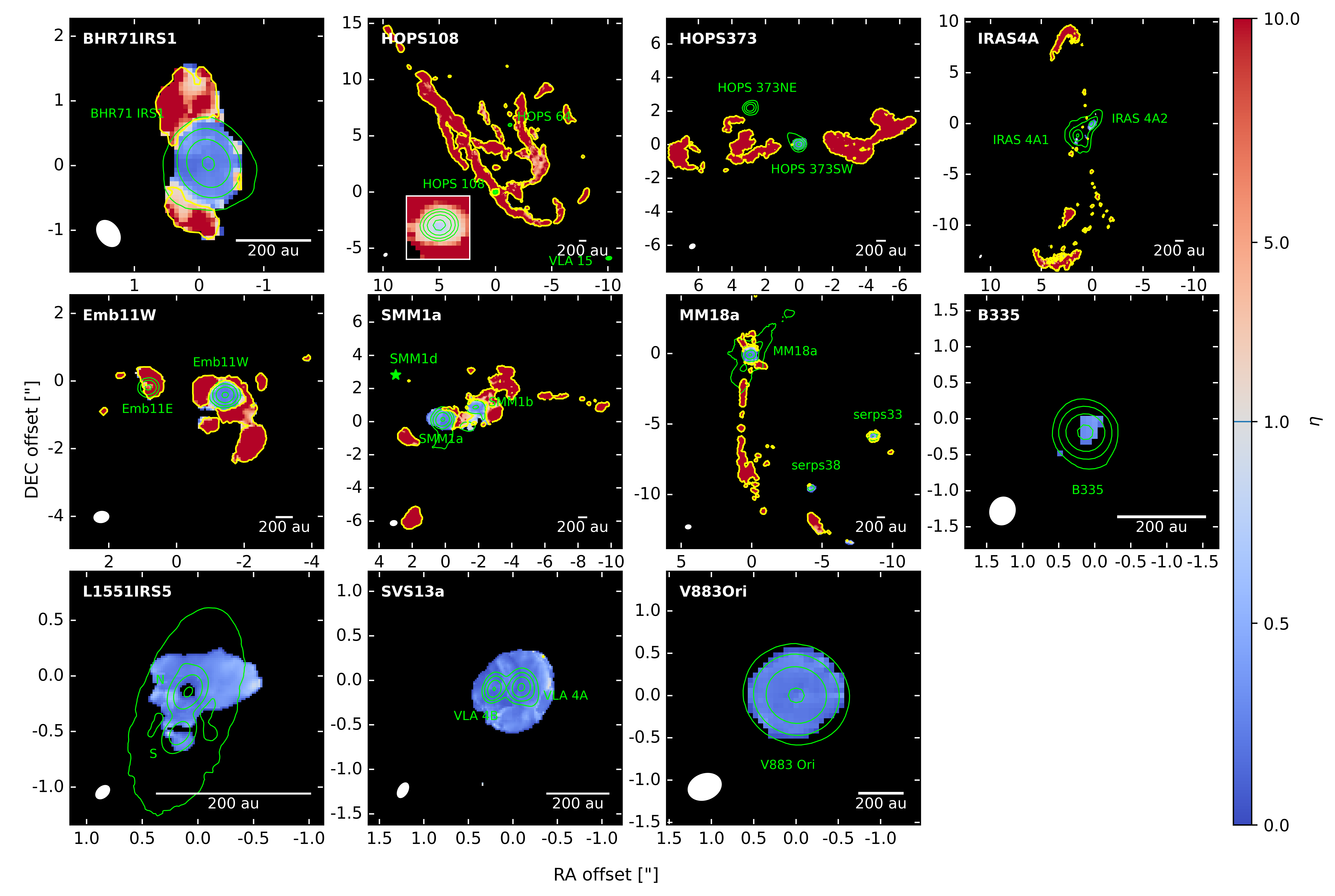}
        \caption{Rotation diagram-based $\eta$ maps of COMPASS sources (Sect. \ref{sec:ratio map}), highlighting regions associated with the 302.9~GHz CH$_3$OH maser emission. In general, the hot corino regions exhibit $\eta$ values below unity, indicating that thermal emission dominates (shown in blue), while the surrounding extended structures often show $\eta$ values greater than one, consistent with maser emission (shown in red). The yellow contours mark regions where the 302.9~GHz data point deviates from the LTE best-fit line to the thermal CH$_3$OH transitions by more than three times the standard deviation in the rotation diagram, indicating that $\eta$ clearly remains greater than unity even after accounting for the scatter. Green contours show the 1 mm continuum, starting at 75$\sigma$ and increasing logarithmically up to 90$\%$ of the peak intensity. SMM1d, one of the outflow-driving sources in the SMM1 field \citep{Hull2017}, is not visible at these contour levels; its position is therefore marked with a green star. The synthesised beam is shown as a white ellipse in the bottom-left corner of each panel. The inset in the HOPS108 panel shows a zoomed-in view of the hot corino of HOPS108.}
        \label{fig:RD ratio map}
    \end{figure*}

In the previous subsection, we presented the results of the rotation diagram analysis at one representative position. Here, we extend this analysis across all spatial pixels where the brightness temperature of the 302.9~GHz CH$_3$OH maser candidate transition exceeds the 4$\sigma$ noise level, to gain insight into the environments in which it preferentially arises.\par

We defined a departure factor `$\eta$' from the rotation diagram analysis as 
\begin{center}
\(
\mathrm{\eta} \equiv \exp\left(\Delta \right)
\),

\(
\Delta \equiv \ln(N_{\rm u}/g_{\rm u})_{\rm measured} - \ln(N_{\rm u}/g_{\rm u})_{\rm LTE\:predicted}
\),
\end{center}
i.e. $\Delta$ is the offset between the observed $\ln(N_{\rm u}/g_{\rm u})$ value of the 302.9~GHz CH$_3$OH maser candidate and the value predicted by the best-fit linear trend under the assumption of LTE (red linear fit in Fig.~\ref{fig:rotation diagram}). This $\eta$ quantifies the degree to which the maser candidate deviates from the thermal population distribution at each pixel.\par

Figure~\ref{fig:RD ratio map} shows the $\eta$ maps, and in the positions where all transitions are optically thin, such as extended regions, the $\eta$ would be unity if the 302.9~GHz transition is purely thermal. In the presence of maser amplification, $\eta$ is expected to exceed unity. However, in the optically thick regime such as hot corinos, purely thermal emission would result in an $\eta$ value below unity (e.g. from B335 to V883Ori in Fig.~\ref{fig:rotation diagram}), and additional non-thermal emission would raise the value towards unity. To assess whether values of $\eta>1$ could arise simply from the intrinsic scatter of the thermal CH$_3$OH lines, we measured the standard deviation of the residuals between the thermal CH$_3$OH data points and the best-fit line shown in the rotation diagrams. The yellow contours in Fig.~\ref{fig:RD ratio map} then mark regions where the 302.9~GHz CH$_3$OH data points are shifted above the best-fit line ($\eta=1$) by more than three times this standard deviation.\par

The 11 central hot corinos targeted in COMPASS show $\eta$ below unity consistent with thermal emission. Other continuum sources in the maps also show thermal emission, including SMM1b \citep{Choi2009}, and the two continuum sources southwest of MM18a (serps33 or CARMA-6 and serps38; \citealt{Plunkett2015, Plunkett2018}), consistent with hot corino origins. The $\eta$ above unity observed towards Emb11E indicates that its 302.9~GHz CH$_3$OH emission arises from maser activity in a shocked gas that is likely located along the line of sight towards Emb11E. This agrees with a COMs survey that reported no hot corino associated with Emb11E \citep{Mart2021}. In addition, in HOPS108 the hot corino itself shows $\eta$ values close to unity across its extent. This implies that maser amplification by shock-excited gas along the line of sight—driven by outflow shocks from the adjacent intermediate-mass protostar HOPS370 and located in front of the hot corino—enhances the line intensity enough to counteract the reduction expected for a thermally optically thick line, thereby producing $\eta$ values close to or even above unity. In most of the extended regions of the seven sources with maser detection, $\eta$ values are higher than one, demonstrating that maser emission is widespread.\par

We also examined the ratio between the full width at half maximum (FWHM) of the 302.9~GHz line and that of the thermal CH$_3$OH lines using the pixel-by-pixel Gaussian fitting results. For the maser-detected sources, the FWHM-ratio distributions are broadly concentrated around values close to unity, with median values typically in the range of $\sim$0.8--0.95. Among the maser-detected sources, HOPS108 shows the smallest median ratio of $\sim$0.76, suggesting that the 302.9~GHz line is relatively narrower there. Considering the velocity resolution of our data ($\sim$0.5~km~s$^{-1}$), however, these modest offsets from unity are difficult to interpret as a robust linewidth difference between the 302.9~GHz line and the thermal CH$_3$OH lines. We therefore conclude that linewidth alone does not provide a robust discriminator between maser and thermal CH$_3$OH emission in the COMPASS sample, and it should be interpreted together with the $\eta$-based non-LTE analysis and the emission morphology.\par

To conclude the analysis of rotation diagrams, the 302.9~GHz CH$_3$OH maser emission is detected towards BHR71IRS1, HOPS108, HOPS373, IRAS4A, Emb11W, SMM1a, and MM18a. The observed maser brightness temperature spans a wide range across the sample: in several sources the 302.9 GHz line is comparable to, weaker than, or only modestly brighter than other thermal CH$_3$OH transitions at the same positions, whereas in a few cases, most notably HOPS108, it is significantly brighter. The linewidths, however, are often comparable to those of the thermal CH$_3$OH lines.\par

Nevertheless, the $\eta$ maps show that the maser intensity exceeds the LTE prediction over large areas (Fig. \ref{fig:RD ratio map}), demonstrating that the 302.9~GHz CH$_3$OH emission is genuinely non-thermal. Importantly, these $\eta$ > 1 regions form extended arcs and filaments that closely follow the bow shocks and outflow cavity walls, whereas the hot corino regions near the continuum peaks remain mostly consistent with thermal emission. Thus, the 302.9~GHz transition traces a spatially extended maser, operating throughout the outflow-driven low-velocity shocks.

\section{Discussion} \label{sec:discussion}
\subsection{Widespread occurrence of class I CH$_3$OH masers in low-mass protostars}
The 302.9~GHz class I CH$_3$OH maser emission is detected in seven out of 11 protostars in the COMPASS sample. Excluding HOPS108, where the emission is associated with the intermediate-mass protostar HOPS370, maser emission is detected in six of the remaining ten low-mass protostars, corresponding to a detection rate of 60$\%$.\par

This detection rate is remarkably high, given that class I CH$_3$OH masers have historically been regarded as rare in low-mass protostellar environments and primarily associated with high-mass star formation (e.g. \citealt{Kalenskii2006, Kalenskii2017}). For example, \citet{Gan2013} surveyed 165 low-mass molecular outflow sources using the 13.7 m Purple Mountain Observatory telescope and detected 95~GHz class I CH$_3$OH masers in only 15 sources (9\%). Similarly, \citet{Kang2013} surveyed 99 low-mass protostars in the Orion molecular cloud complex with the 21 m KVN telescope, targeting 44, 95, and 133~GHz CH$_3$OH maser transitions, and detected masers in only four sources (4\%). More recently, \citet{Humire2024} used the 12 m Atacama Pathfinder Experiment (APEX) telescope to detect 326.9~GHz class I CH$_3$OH $10_{1,10}-9_{-0,9}$E maser emission in four out of 19 low-mass protostars, corresponding to a detection rate of 21\%.\par

\begin{figure}[t]
        \centering
        \includegraphics[trim=0cm 0cm 0cm 0cm, clip=true,width=0.49\textwidth]{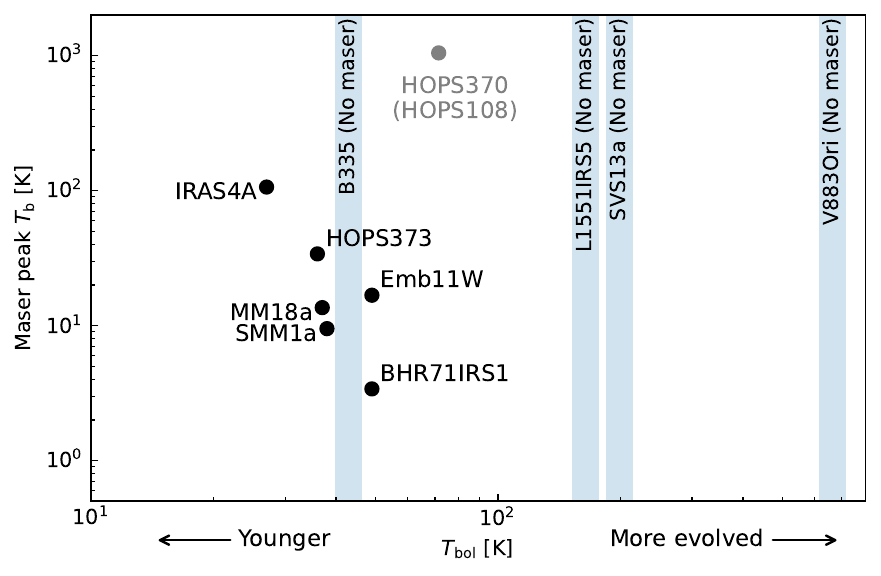}
        \caption{Maser peak $T_{\rm b}$ (primary-beam corrected) as a function of $T_{\rm bol}$ for the maser-detected sources. Sources with no detected maser emission are marked by vertical shaded bands at their respective $T_{\rm bol}$ values. The y-axis represents the $T_{\rm b}$ value at the brightest pixel in the 302.9~GHz CH$_3$OH maser candidate map (red cross in Fig.~\ref{fig:h2co_methanol mom8 map}). The $T_{\rm bol}$ values are from \citet{jorgensen25compass}. HOPS370, although not a COMPASS target, is included because it is the primary source influencing the HOPS108 region. Its $T_{\rm bol}$ value is from \citet{Furlan2016}. HOPS370 is marked in grey in the figure to distinguish it as an intermediate-mass protostar. Note that the class boundaries of protostars follow \citet{Chen1995}: Class 0 (70 K> $T_{\rm bol}$ > 20 K) and Class I (650 K > $T_{\rm bol}$ > 70 K).}
        \label{fig:Tbolvsmaser}
    \end{figure}

These low detection rates likely reflect observational limitations rather than a true absence of masers. Many of the COMPASS detections exhibit weak peak brightness temperatures (Fig.~\ref{fig:Tbolvsmaser}), which would have been difficult to identify in single-dish observations. In particular, single-dish surveys suffer from beam dilution, which suppresses the signals from compact and faint masers.

\subsection{Conditions favouring the class I CH$_3$OH maser excitation}
While the high detection rate suggests that class I CH$_3$OH masers are more common in low-mass protostars than previously thought, their occurrence is not uniform across all sources. As shown in Fig. \ref{fig:Tbolvsmaser}, the six detections are associated with Class 0 protostars, whereas three of the four non-detections correspond to more evolved Class I or Class I/II systems; the isolated core B335 is the only non-detection in a Class 0 source. Furthermore, among the six low-mass maser-detected sources, the maser peak $T_{\rm b}$ appears to decline with increasing $T_{\rm bol}$, although this trend should be regarded as tentative given the small sample size. We note that `class I' in lowercase refers to the maser excitation mechanism (collisionally pumped) and should not be confused with the protostellar evolutionary classification denoted by uppercase `Class I'.\par

Although both Class 0 and I objects are embedded protostars, Class I systems have lower envelope masses as accretion progresses (e.g. \citealt{Lada1987, Andre1993, Tobin2024}). The reduced envelope mass in Class I sources likely results in less surrounding material for outflows to interact with, thereby limiting the amount of CH$_3$OH gas that can be overpopulated.\par

The bolometric temperature of HOPS370 ($T_{\rm bol}=71.5$ K) places it near the boundary between Class 0 and Class I. The outflow driven by HOPS370 actively interacts with the FIR 4 region of OMC-2, where HOPS108 resides. FIR 4 hosts the brightest 1.3 mm continuum emission in OMC-2 and contains dense dust condensations \citep{Sato2023}. Regardless of its intermediate mass, the detection of the 302.9~GHz maser emission towards the more evolved HOPS370 may be facilitated by the dense surrounding material in the FIR 4 environment, in contrast to the other three low-mass Class I sources (L1551IRS5, SVS13a, and V883Ori), which show no maser detections.\par

Our analysis further shows that maser emission is absent in hot corinos even though strong 302.9~GHz CH$_3$OH emission is present (Fig.~\ref{fig:RD ratio map}). Since hot corinos are the densest regions compared to extended, ambient materials, any shock-induced excitation is likely to be rapidly thermalised.

 Another factor that favours excitation of the 302.9~GHz CH$_3$OH maser is the presence of a powerful outflow. As shown in Fig. \ref{fig:Tbolvsmaser}, the masers around HOPS108 (or HOPS370) are more than an order of magnitude brighter than those in other sources. This is consistent with previous findings that the intermediate-mass protostar HOPS370 drives an unusually strong outflow, with a mass-loss rate more than ten times higher than that of typical low-mass protostars \citep{Nisini2015, Gonz2016}.\par
 
Altogether, our results suggest that the newly discovered 302.9~GHz class I CH$_3$OH maser is most efficiently excited in environments where both strong outflow-driven shocks (such as HOPS370) and abundant surrounding material are present (as in earlier evolutionary stages). The remarkably high detection rate towards low-mass protostars revealed by COMPASS, enabled by its high sensitivity and wide unbiased spectral coverage, demonstrates that class I CH$_3$OH masers are far more prevalent in low-mass star-forming regions than previously recognised. These findings have important implications for future maser surveys of low-mass protostellar environments.

\subsection{Complex chemistry at shocks traced by the masers} \label{sec:CH3CHO}

The fact that the shock-induced class I 302.9~GHz CH$_3$OH maser is very commonly observed in low-mass protostars suggests that the class I CH$_3$OH maser has the potential to serve as a major shock tracer in low-mass protostellar systems. To explore this possibility, we investigate the connection between the maser and the shock chemistry of COMs in this section.

    \begin{figure*}[t]
        \centering
        \includegraphics[trim=0cm 0cm 0cm 0cm, clip=true,width=0.9\textwidth]{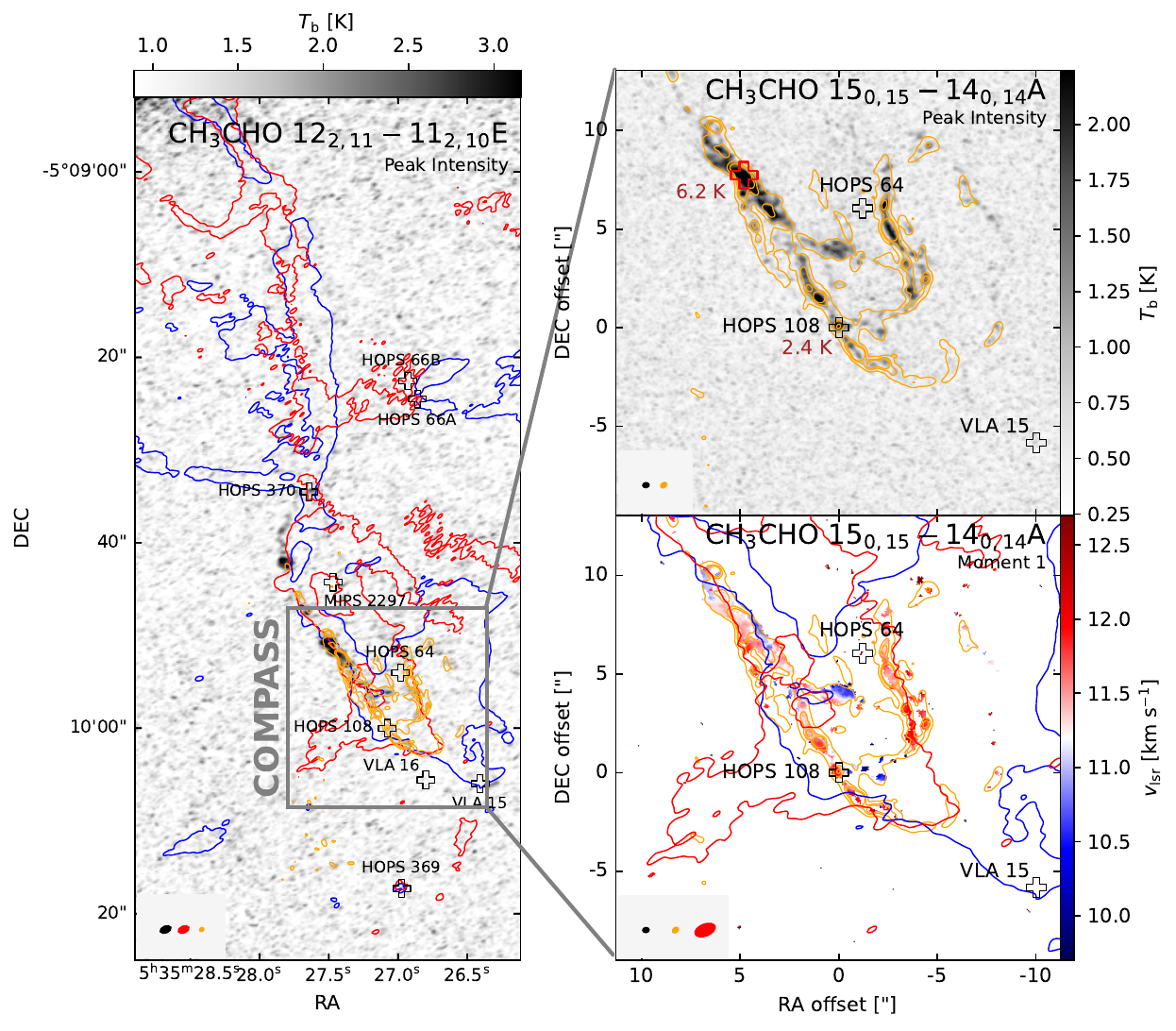}
        \caption{Maps of CH$_3$CHO in the OMC-2 FIR 3 (location of HOPS370) and FIR 4 (location of HOPS108) regions in a wide field of view using ALMA archival Band 6 observations (project 2017.1.01353.S, left panel) and in the FIR 4 region in the COMPASS field of view (right panels). Left panel (archival data): Peak $T_{\rm b}$ map of CH$_3$CHO $12_{2,11}-11_{2,10}$E ($E_{\rm u}=81.1$~K, log$_{10}(A_{\rm ul}(\rm s^{-1}))=-3.38$, 230.316~GHz). Top-right panel (COMPASS data): Peak $T_{\rm b}$ map of CH$_3$CHO $15_{0,15}-14_{0,14}$A ($E_{\rm u}=109.2$~K, log$_{10}(A_{\rm ul}(\rm s^{-1}))=-3.10$, 281.127~GHz). Bottom-right panel (COMPASS data): Intensity-weighted velocity map of CH$_3$CHO $15_{0,15}-14_{0,14}$A ($E_{\rm u}=109.2$~K, log$_{10}(A_{\rm ul}(\rm s^{-1}))=-3.10$, 281.127~GHz) with respect to HOPS370's velocity (v$_{\rm LSR}$ = 11.2~km s$^{-1}$; \citealt{Tobin2019}). The $T_{\rm b}$ values written in the upper panel are the peak $T_{\rm b}$ of CH$_3$CHO at the 302.9~GHz CH$_3$OH maser peak position (red cross in Fig. \ref{fig:h2co_methanol mom8 map}) and the hot corino region of HOPS108. Common contours: The continuum sources identified from previous studies are indicated with black crosses. The orange contours represent the peak $T_{\rm b}$ maps of the 302.9~GHz CH$_3$OH maser emission shown in Fig. \ref{fig:h2co_methanol mom8 map}. The blue and red contours show integrated $T_{\rm b}$ maps of the blue- and redshifted $^{12}$CO $J=2-1$ transition. The emission was integrated over velocity ranges that are $4-6$~km s$^{-1}$ blueshifted and $4-6$~km s$^{-1}$ redshifted with respect to HOPS370. For simplicity, we only show one contour level at $3\sigma$. The black, red, and orange ellipses in the bottom-left corner indicate the beams of CH$_3$CHO, $^{12}$CO $J=2-1$, and the 302.9~GHz CH$_3$OH maser, respectively.}
        \label{fig:ch3cho large map}
    \end{figure*}

\subsubsection{Detection of acetaldehyde (CH$_3$CHO) along the class I 302.9~GHz CH$_3$OH masing components}

We analysed the unbiased spectral survey data for the seven CH$_3$OH maser-detected COMPASS targets to identify which COMs are preferentially present under the conditions in which the CH$_3$OH maser is detected. Among the seven maser sources, HOPS108 is the only source in which we find a molecular species, CH$_3$CHO, whose emission morphology closely matches that of the 302.9 GHz CH$_3$OH maser. Accordingly, this subsection does not address shock chemistry in low-mass protostars more broadly but instead focuses specifically on the shock chemistry associated with the outflow driven by the intermediate-mass protostar HOPS370.\par

In the spectra extracted from the brightest CH$_3$OH maser pixel (red cross in the panel of HOPS108 in Fig. \ref{fig:h2co_methanol mom8 map}), we also detect, for example, formamide (NH$_2$CHO), methyl cyanide (CH$_3$CN), dimethyl ether (CH$_3$OCH$_3$), methyl formate (CH$_3$OCHO), and propyne (CH$_3$CCH). However, CH$_3$CHO is unique in showing brighter, extended emission than at the central hot corino position, mirroring the maser behaviour. We therefore focus on CH$_3$CHO in HOPS108 in the following discussion.\par

To investigate the interaction between the outflow shocks from HOPS370 and the surrounding material in OMC-2 FIR 4, we utilised ALMA archival Band 6 observations covering the FIR 3 and FIR 4 region (project 2017.1.01353.S; P.I.: Satoko Takahashi; see \citealt{Sato2023}). These data provide a significantly larger field of view than the COMPASS observations. We extracted the pipeline imaging products (FITS cube) from the ALMA Science Archive for two transitions: $^{12}$CO $J=2-1$ (230.538 GHz) and CH$_{3}$CHO $12_{2,11}-11_{2,10}$E (230.316 GHz), as the CH$_3$CHO transition lies within the $^{12}$CO spectral window. The synthesised beam is $1\farcs16 \times 0\farcs67$, with a velocity resolution of $0.3$~km s$^{-1}$ and a sensitivity of $16.8$ mJy beam$^{-1}$.\par

The left panel of Fig. \ref{fig:ch3cho large map} shows the peak brightness temperature map of the CH$_3$CHO $12_{2,11}-11_{2,10}$E transition, and the integrated brightness temperature contours of blue- and redshifted $^{12}$CO $J=2-1$ emission. The orange contours of the 302.9~GHz CH$_3$OH maser peak brightness temperature appear exclusively within the grey rectangle labelled ‘COMPASS’. The CH$_3$CHO emission is confined to the eastern edge of the southern, redshifted CO outflow lobe of HOPS370; no CH$_3$CHO is detected along the northern lobes. Its brightness is on the order of a few kelvin and is higher along the outflow lobe than at the hot corino of HOPS370 and HOPS108. The right panels display the COMPASS observations, enabling a detailed comparison of the spatial and kinematic characteristics of the 302.9~GHz CH$_3$OH maser, the CO outflows, and the CH$_3$CHO emission. The CH$_3$CHO line shown is the $15_{0,15}-14_{0,14}$A transition, the brightest isolated CH$_3$CHO transition covered in the COMPASS spectral survey at the red cross position in Fig.~\ref{fig:h2co_methanol mom8 map}. The CH$_3$CHO map in the right panel reveals additional emission that is not seen in the left panel map, which we attribute to the higher sensitivity of the COMPASS data relative to the ALMA archival data.\par

The peak brightness temperature map (upper right) shows that the CH$_3$OH maser emission and CH$_3$CHO emission have nearly coincident spatial distributions: the maser peak coincides with the CH$_3$CHO peak (6.2~K). The CH$_3$CHO velocity field (lower right) is predominantly redshifted by 1~km s$^{-1}$ relative to HOPS370 (v$_{\rm LSR}$ = 11.2~km s$^{-1}$), consistent with a low-velocity shock. Adopting an inclination of 74$^{\circ}$ for the outflow of HOPS370 \citep{Tobin2020_2}, an observed 1~km s$^{-1}$ corresponds to 3.6~km s$^{-1}$ after deprojection.

\subsubsection{Gas-phase shock chemistry of CH$_3$CHO and rotation diagram analysis}

    \begin{figure*}[t]
        \centering
        \includegraphics[trim=0cm 0cm 0cm 0cm, clip=true,width=0.9\textwidth]{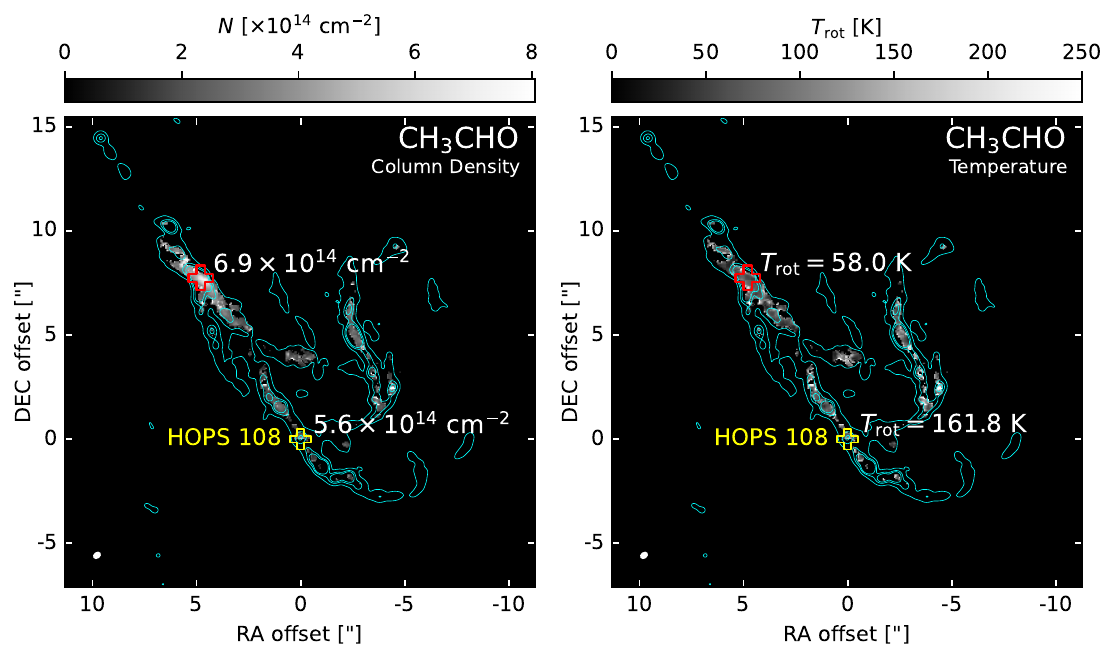}
        \caption{Column density (left) and rotation temperature (right) maps of CH$_3$CHO around HOPS108 and the OMC-2 FIR 4 region based on the rotation diagram analysis. The column density and rotation temperature at the 302.9~GHz CH$_3$OH maser peak position (red cross in Fig. \ref{fig:h2co_methanol mom8 map}) and in the hot corino region of HOPS108 in the centre of the map are written in white. The cyan contours represent the peak $T_{\rm b}$ maps of the 302.9~GHz CH$_3$OH maser emission shown in Fig. \ref{fig:h2co_methanol mom8 map}. The white ellipses in the bottom-left corners denote the beam size of CH$_3$CHO $15_{0,15}-14_{0,14}$A.}
        \label{fig:ch3cho RD map}
    \end{figure*}

The fact that CH$_3$CHO is brightest along the outflow lobe suggests rapid shock-driven formation. Protostellar shock chemistry of COMs is initiated by non-thermal desorption via grain sputtering (i.e. dust–gas collisions), followed by gas-phase reactions at elevated temperatures \citep{Burkhardt2019}. In particular, for COMs in ice mantles to desorb into the gas phase without being destroyed, non-dissociative shocks (e.g. C-type shocks) are required \citep[e.g.][]{Dishoeck1998}.\par

Shock-induced chemistry is species dependent. Modelling along outflow-cavity walls and for C-shocks shows that CH$_3$CHO undergoes strong gas-phase enhancement once precursors are injected by sputtering \citep{Burkhardt2019}, or enhanced UV irradiation \citep{Drozdovskaya2015}; an efficient route is CH$_3$CH$_2$ + O $\rightarrow$ CH$_3$CHO + H \citep{Charnley2004}. Enhanced gas-phase CH$_3$CHO has been observed in several outflow cavities, including L1157-B0/B1/B2 \citep{Codella2015, Codella2017, Codella2020, Lopez2024}, IRAS4A \citep{Simone2020}, and tentatively in a CH$_3$OH-rich component located between the protostellar members of the proto-triple system HOPS 288 \citep{Hsu2026}.

To test whether shock chemistry operates in the extended CH$_3$CHO structure, we perform a pixel-by-pixel rotation diagram analysis. Figure~\ref{fig:ch3cho RD map} shows the resulting CH$_3$CHO column density and rotation temperature maps (see Sect. B for methods). Column densities are generally higher where both the CH$_3$OH maser and CH$_3$CHO are bright. The value at the maser peak (6.9$\times$10$^{14}$~cm$^{-2}$) is slightly higher than that at the HOPS108 hot corino (5.6$\times$10$^{14}$~cm$^{-2}$). The rotation temperature map indicates $T_{\rm rot}\!\approx\!162$~K at the hot corino, typical for such environments, and a lower temperature of $T_{\rm rot}\!\approx\!58$~K at the red cross in Fig.~\ref{fig:h2co_methanol mom8 map}.\par

Beyond CH$_3$CHO, two other COMs (CH$_3$OCHO and NH$_2$CHO) are also expected to be significantly enhanced in post-shock gas \citep{Burkhardt2019}, with models predicting lower CH$_3$CHO abundances than these two species. Nevertheless, along the shocked region around FIR 4, neither CH$_3$OCHO nor NH$_2$CHO shows extended emission comparable to CH$_3$CHO; outside the hot corino, 
their peak brightness temperatures are only about three to five times the noise level near the brightest 302.9~GHz CH$_3$OH maser spot. This discrepancy may reflect differences in pre-shock ice composition or post-shock chemical timescales \citep{Burkhardt2019}. Further observations and modelling are needed to better quantify the relationship between the 302.9~GHz CH$_3$OH maser emission and the shock chemistry of complex organics.\par

In summary, our analysis suggests that in environments that produce bright class I 302.9~GHz CH$_3$OH masers, CH$_3$CHO can be formed efficiently through shock-driven chemistry. However, the present analysis is limited to the HOPS108/HOPS370 region, where the shock is driven by the intermediate-mass protostar HOPS370. To determine whether a similar relationship also holds for the other low-mass protostars with the 302.9~GHz maser, more sensitive observations will be required.

\section{Summary} \label{sec:summary}
We have presented the serendipitous discovery of the millimetre/submillimetre class I CH$_3$OH maser ($12_{0,12}-11_{1,11}$A, $E_{\rm u}=180.9$~K, 302.913~GHz) in low-mass protostellar environments by examining the unbiased spectral survey data (279.0–311.7~GHz) of the ALMA Large Program COMPASS. We also explored the connection of this newly detected maser to the COM acetaldehyde (CH$_{3}$CHO) in the context of shock chemistry, specifically for HOPS108. Our main findings are as follows.

\begin{enumerate}
    \item The 302.9~GHz CH$_3$OH maser is detected in seven out of 11 COMPASS targets ($\sim$64\%). However, one of these detections, towards HOPS108, is likely associated with the powerful outflow driven by the nearby intermediate-mass protostar HOPS370 rather than with the low-mass protostar itself. Excluding this case, the maser is detected in six out of ten low-mass protostars ($60\%$). This detection rate indicates that class I CH$_3$OH masers are more common in low-mass protostars than previously appreciated.
    
    \item While the linewidths are generally comparable to those of the thermal CH$_3$OH lines, the observed maser intensity shows source-to-source variation, ranging from thermal levels in several sources to exceptionally bright emission in the region around HOPS108. The dense FIR 4 environment around HOPS108, interacting with strong outflows driven by HOPS370, hosts the brightest maser.

    \item The occurrence of the 302.9~GHz maser correlates with the early evolutionary stage (Class 0) and dense environments with substantial outflows. Non-detections are concentrated among the more evolved Class I sources with less ambient material, although the maser is also absent in the Class 0 isolated core B335. 

    \item We introduced a pixel-based diagnostic that compares the observed $\ln(N_{\rm u}/g_{\rm u})$ of the 302.9~GHz line to the LTE prediction obtained from fitting other CH$_3$OH transitions. This diagnostic cleanly separated thermal from non-thermal emission, and it revealed that the 302.9~GHz CH$_3$OH masing regions trace bow shocks and outflow cavity walls, while hot corinos appear as thermally dominated.

    \item Around HOPS108, in the region affected by the strong outflow from the intermediate-mass protostar HOPS370, CH$_3$CHO emission closely matches the spatial distribution of the 302.9~GHz CH$_3$OH maser. Rotation diagram mapping indicates a slightly higher CH$_3$CHO column density at the maser peaks than towards the hot corino of HOPS108, consistent with CH$_3$CHO enhancement driven by shock-induced sputtering or enhanced UV irradiation, followed by efficient warm gas-phase formation in low-velocity shocks.
\end{enumerate}
    These results establish the 302.9~GHz CH$_3$OH transition as both a sensitive tracer of outflow–shock interfaces in low-mass protostars and a signpost of active complex organic processing. Unbiased spectral surveys with interferometers can uncover weak masers, opening a new window to the physical and chemical impact of protostellar shocks on their extended environments.

\section*{Data availability}
The additional figures and table are available in the Zenodo repository: \url{https://doi.org/10.5281/zenodo.21637507}.

\begin{acknowledgements}
This work is based on the following ALMA data: ADS/JAO.ALMA\#2022.1.00316.L. ALMA is a partnership of ESO (representing its member states), NSF (USA), and NINS (Japan), together with NRC (Canada), NSTC and ASIAA (Taiwan), and KASI (Republic of Korea), in cooperation with the Republic of Chile. The Joint ALMA Observatory is operated by ESO, AUI/NRAO, and NAOJ. The project leading to this publication has received support from ORP, that is funded by the European Union’s Horizon 2020 research and innovation programme under grant agreement No 101004719 [ORP]. JEL, JHJ, HSY, and CHK were supported by the NRF grant funded by the Korean government (MSIT) (grant numbers RS-2024-00416859 and RS-2026-25490557). JHJ was also supported by Basic Science Research Program through the National Research Foundation of Korea (NRF) funded by the Ministry of Education (grant number RS-2025-25402622). AC and FCSM received financial support from the European Research Council (ERC) under the European Union’s Horizon 2020 research and innovation programme (ERC Starting Grant “Chemtrip”, grant agreement No 949278). SM acknowledges support from the Thematic Action “Physique et Chimie du Milieu Interstellaire” (PCMI) of INSU Programme National “Astro”, with contributions from CNRS Physique \& CNRS Chimie, CEA, and CNES. DH is supported by a Center for Informatics and Computation in Astronomy (CICA) grant and grant number 110J0353I9 from the Ministry of Education of Taiwan. DH also acknowledges support from the National Science and Technology Council, Taiwan (Grant NSTC111-2112-M-007-014-MY3, NSTC113-2639-M-A49-002-ASP, and NSTC113-2112-M-007-027). AK was supported by the NKFIH NKKP grant ADVANCED 149943 and the NKFIH excellence grant TKP2021-NKTA-64. Project no.149943 has been implemented with the support provided by the Ministry of Culture and Innovation of Hungary from the National Research, Development and Innovation Fund, financed under the NKKP ADVANCED funding scheme. This publication is based upon work from COST Action PLANETS CA22133, supported by COST (European Cooperation in Science and Technology). PN acknowledges support from the ESO and IAU Gruber Foundation Fellowship programs. SS acknowledges the support of the Max Planck Society. YLY acknowledges support from a Grant-in-Aid from the Ministry of Education, Culture, Sports, Science, and Technology of Japan (20H05845, 20H05844), and a pioneering project in RIKEN (Evolution of Matter in the Universe). SZ thanks RIKEN Special Postdoctoral Researcher Program (Fellowships) for financial support.\\
Facility: ALMA.\\
Software: Numpy \citep{Harris2020}, Scipy \citep{Virtanen2020}, Astropy \citep{astropy2013, Price-Whelan_2018, astropy2022}, Pandas \citep{McKinney2010}, Matplotlib \citep{matplotlib}, XCLASS \citep{Moller17}
\end{acknowledgements}

\bibliographystyle{aa}
\bibliography{aa58734-25}

\end{document}